\documentclass[10pt,a4paper]{article}
\usepackage[a4paper, left=2cm, right=2cm, top=1cm]{geometry}

\usepackage[utf8]{inputenc}
\usepackage{amsmath, amsthm, amssymb, amsfonts}     
\usepackage{mathrsfs}                               
\usepackage{accents}                                
\usepackage{natbib}                                 
\usepackage{hyperref}                               
\usepackage{doi}                                    
\usepackage{authblk}                                
\usepackage{siunitx}                                
\usepackage{booktabs}                               
\usepackage{ulem}                                   
\usepackage{nomencl}                                
\makenomenclature                                   
\usepackage[linesnumbered,ruled,vlined]{algorithm2e}
\SetCommentSty{mycommfont}
\usepackage{subfigure}

\usepackage{bm}                     

\usepackage{graphicx}
\usepackage{tikz}
\usepackage{pgfplots}

\usetikzlibrary{decorations.pathmorphing,arrows,intersections,arrows.meta,angles,quotes}

\pgfplotsset{
	kurze Legende/.style={
		legend image code/.code={
			\draw[##1,mark repeat=2,line width=0.6pt]
			plot coordinates {
				(0cm,0cm)
				(0.3cm,0cm)
			};
		}
	}
}

\pgfplotsset{
	compat = newest,
	scale only axis, 
	max space between ticks = 50pt,
	ticklabel style = {font=\footnotesize},
	legend style =  {font=\footnotesize},
	grid = major,
	grid style = {dotted},
	legend columns=1, 
	xtick pos=left,
	ytick pos=left
}

\pgfplotsset{select coords between index/.style 2 args={
		x filter/.code={
			\ifnum\coordindex<#1\fi
			\ifnum\coordindex>#2\fi
		}
}}

\definecolor{color1}{HTML}{0060AD} 
\definecolor{color2}{HTML}{FF4500} 
\definecolor{color3}{HTML}{FFA500} 
\definecolor{color4}{HTML}{006400} 
\definecolor{color5}{HTML}{9400D3} 
\definecolor{color6}{HTML}{800000} 
\definecolor{color7}{HTML}{000000} 
\definecolor{color8}{HTML}{0000FF} 
\definecolor{color9}{HTML}{FF0000} 
\definecolor{mycolor_blue}{RGB}{66,124,161}
\definecolor{mycolor_grey}{RGB}{198,198,198} 

\tikzstyle{line1} = [color=color7,semithick] 
\tikzstyle{line2} = [color=color2,densely dotted,semithick]
\tikzstyle{line3} = [color=color1,densely dashed,semithick]
\tikzstyle{line4} = [color=color5,dash dot,semithick]
\tikzstyle{line5} = [color=color4,dash dot dot,semithick]
\tikzstyle{line6} = [color=color6,semithick]

\tikzstyle{mark1} = [color=color7,mark=x,mark size=2pt,mark options=solid,semithick] 
\tikzstyle{mark2} = [color=color2,mark=o,mark size=2pt,mark options=solid,semithick]
\tikzstyle{mark3} = [color=color1,mark=*,mark size=2pt,mark options=solid,semithick]
\tikzstyle{mark4} = [color=color5,mark=triangle,mark size=2pt,mark options=solid,semithick]
\tikzstyle{mark5} = [color=color4,mark=square,mark size=2pt,mark options=solid,semithick]
\tikzstyle{mark6} = [color=color7,mark=o,mark size=2pt,mark options=solid,semithick]
\tikzstyle{mark7} = [color=color7,mark=*,mark size=2pt,mark options=solid,semithick]
\tikzstyle{mark8} = [color=color7,mark=triangle,mark size=2pt,mark options=solid,semithick]

\usetikzlibrary{external}
\title{Adjoint Sensitivity Maps for Passive Flow Control Around Rotating Circular Cylinders Across a Wide Operating Envelope}

\author[1,2]{Niklas K\"uhl\thanks{kuehl@hsva.de}}

\affil[1]{Hamburg Ship Model Basin, Bramfelder Strasse 164, D-22305 Hamburg, Germany}
\affil[2]{University of Rostock, Chair of Computational Methods for Fluid Dynamics, Albert-Einstein-Straße 2, D-18059 Rostock, Germany}

\begin{document}

\providetoggle{tikzExternal}
\settoggle{tikzExternal}{true}
\settoggle{tikzExternal}{false}

\maketitle

\begin{abstract}
Rotating circular cylinders are employed in a variety of engineering applications, one prominent example being Flettner rotors for wind-assisted ship propulsion. Besides optimizing the aerodynamic performance of the cylinder itself, passive flow-control devices placed in its vicinity offer additional potential for manipulating the resulting aerodynamic forces. The present work introduces a topology-based adjoint sensitivity analysis for rotating circular cylinders over a wide operating envelope covering Reynolds numbers from $10^1$ to $10^7$ and spinning ratios between $0$ and $2\pi$. Local sensitivity fields associated with drag, lift, and torque are derived using a porous-medium formulation and validated by dedicated forward simulations employing both distributed Darcy-type source terms and a sensitivity-informed passive flow-control structure. Particular emphasis is placed on the combined sign distribution of the drag and lift sensitivities, yielding intuitive design maps that directly identify regions where local momentum extraction simultaneously improves or deteriorates both objectives. A systematic investigation of the resulting sensitivity spectra reveals that the large-scale topology of the sensitivity fields is governed primarily by the spinning ratio, whereas the influence of the Reynolds number remains comparatively weak over large parts of the investigated operating envelope. The resulting sensitivity atlas provides practical design guidance for passive flow-control concepts and demonstrates that robust solutions may exist over moderate operating ranges.
\end{abstract}

\begin{flushleft}
\end{flushleft}

\section{Introduction}
\label{sec:introduction}

Rotating circular cylinders in a homogeneous flow have experienced renewed interest in recent years, for example through their application as Flettner rotors for wind-assisted ship propulsion \cite{traut2014propulsive, lu2020ship}. By exploiting the Magnus effect, these devices generate large transverse forces that can reduce fuel consumption and greenhouse-gas emissions \cite{seifert2012review, traut2014propulsive}. Consequently, a growing body of research has addressed the aerodynamic characteristics of rotating cylinders over a wide range of Reynolds numbers and spinning ratios using experimental investigations as well as numerical simulations \cite{seifert2012review, mittal2003flow, kang1999laminar}. Particular attention has been devoted to the prediction of drag, lift, and power consumption, which ultimately determine the energetic performance of Flettner-assisted propulsion systems \cite{lu2020ship}.

Besides improving the aerodynamic performance of the rotating cylinder itself, increasing attention has recently been paid to passive flow-control concepts and additional structures placed in the vicinity of the cylinder \cite{choi2008control, zdravkovich1997flow}. Such devices may reduce drag, manipulate the direction of the resulting aerodynamic force, suppress undesired flow features, or improve the overall lift-to-drag ratio \cite{strykowski1990formation, tsutsui2002drag}. However, the design of these structures typically relies on engineering intuition, parameter studies, or computationally demanding optimization procedures, making the systematic identification of favourable locations a challenging task.

Adjoint methods provide an attractive alternative for this purpose \cite{giles2000introduction, jameson1988aerodynamic}. While originally developed for gradient-based optimization, adjoint formulations simultaneously provide detailed sensitivity information describing how local modifications of the governing equations influence a prescribed objective function \cite{giles2000introduction, dilgen2018topology}. In particular, topology-based adjoint methods employing porous-medium formulations enable the identification of favourable and unfavourable regions for passive flow-control concepts without requiring a predefined geometry parameterization \cite{borrvall2003topology, gersborg2005topology, othmer2008continuous}. Beyond optimization, the resulting sensitivity fields therefore constitute a powerful physical analysis tool for interpreting the interaction between the flow field and different aerodynamic objectives.

Although adjoint methods have become well established for a wide variety of aerodynamic applications \cite{othmer2008continuous}, their application to rotating circular cylinders remains limited. Existing studies generally focus on individual operating conditions, e.g., \cite{kuhl2025adjoint}, or specific optimization studies, whereas a systematic investigation of the resulting sensitivity fields over the practically relevant operating envelope has not yet been reported. Consequently, it remains unclear to what extent the sensitivity distributions depend on Reynolds number and spinning ratio and whether robust passive flow-control concepts can be identified that remain effective over a range of operating conditions.

The present work addresses this gap by introducing a topology-based adjoint sensitivity analysis for rotating circular cylinders over Reynolds numbers ranging from $10^1$ to $10^7$ and spinning ratios between $0$ and $2\pi$. Local sensitivity fields associated with drag, lift, and torque are derived using a porous-medium formulation and validated by dedicated forward simulations employing both distributed Darcy-type source terms and a sensitivity-informed passive flow-control structure. Subsequently, a comprehensive sensitivity atlas covering the complete operating envelope is established and interpreted with respect to its physical characteristics and engineering implications. Particular emphasis is placed on the combined sign distribution of the drag and lift sensitivities, resulting in intuitive design maps that directly identify regions where both objectives are simultaneously improved, simultaneously deteriorated, or provide conflicting guidance. To the best of the authors' knowledge, this work presents the first comprehensive adjoint sensitivity atlas for rotating circular cylinders covering seven orders of magnitude in Reynolds number together with the practically relevant range of spinning ratios.

The remainder of this paper is organized as follows. Section~\ref{sec:numerical_method} introduces the governing equations, adjoint formulation, numerical methodology, and uncertainty assessment. The validation studies are presented in Section~\ref{sec:validation}, followed by the systematic analysis of the resulting sensitivity spectra and their engineering implications in Sec. \ref{sec:sensitivity_spectra}. Finally, the main conclusions and directions for future work are summarized.

\section{Numerical Method}
\label{sec:numerical_method}

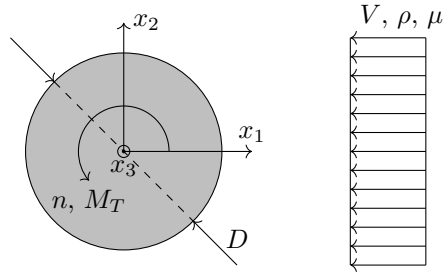
\begin{figure}[!ht]
    \centering
    \iftoggle{tikzExternal}{
    \begin{tikzpicture}

\draw[fill=gray!50] (0,0) circle (1.3); 

\draw[->] (0.6,0) arc (0:220:0.6) node[anchor=north] {$n$, $M_T$}; 

\draw[->] (4,1.50) -- (3,1.50) node[anchor=south west] {$V$, $\rho$, $\mu$};  
\draw[->] (4,1.25) -- (3,1.25);
\draw[->] (4,1.00) -- (3,1.00);
\draw[->] (4,0.75) -- (3,0.75);
\draw[->] (4,0.50) -- (3,0.50);
\draw[->] (4,0.25) -- (3,0.25);
\draw[->] (4,0.00) -- (3,0.00);
\draw[->] (4,-0.25) -- (3,-0.25);
\draw[->] (4,-0.50) -- (3,-0.50);
\draw[->] (4,-0.75) -- (3,-0.75);
\draw[->] (4,-1.00) -- (3,-1.00);
\draw[->] (4,-1.25) -- (3,-1.25);
\draw[->] (4,-1.50) -- (3,-1.50);
\draw[-] (4,-1.50) -- (4,1.50);
\draw[-] (3,-1.50) -- (3,1.50);

\draw[->] (0,0) -- (1.7,0) node[anchor=south] {$x_1$};
\draw[->] (0,0) -- (0,1.7) node[anchor=west] {$x_2$};

\draw (0,0) circle (0.08);
\fill (0,0) circle (0.025);
\node[anchor=north] at (0.0,0.0) {$x_3$};

\draw[dashed] (-0.91923881553, 0.91923881553) -- (0.91923881553, -0.91923881553);
\draw[->] (-1.5,1.5) -- (-0.91923881553, 0.91923881553);
\draw[<-] (0.91923881553, -0.91923881553) -- (1.5,-1.5) node[anchor=south, yshift=3pt] {$D$};

\end{tikzpicture}}{
    \includegraphics{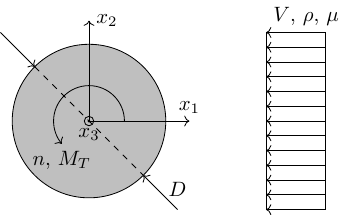}}
    \caption{Two-dimensional rotating-cylinder configuration and definition of the coordinate system.}
    \label{fig:rotating_cylinder_2D}
\end{figure}

The present study investigates sensitivity-based design regions around a rotating cylinder over a wide range of Reynolds numbers and spinning ratios. The employed methodology follows the adjoint-based topological sensitivity framework introduced in the authors' previous work (\cite{kuhl2025adjoint}) and is therefore only summarized briefly in the following. Particular emphasis is placed on the influence of the operating point on the resulting sensitivity distributions and their implications for passive flow-control devices and surrounding structures. Figure \ref{fig:rotating_cylinder_2D} shows the considered two-dimensional cylinder configuration, including the coordinate system, cylinder diameter $D$, free-stream velocity $V$, and rotational frequency $n$, from which the governing Reynolds number and spinning ratio are derived.

The flow field is governed by the incompressible Navier--Stokes equations augmented by a Darcy-type momentum-source term,
\begin{align}
-\frac{\partial v_k}{\partial x_k}
&=
0
\, ,
\\
\frac{\partial \rho v_i}{\partial t}
+
\frac{\partial v_k \rho v_i}{\partial x_k}
+
\frac{\partial}{\partial x_k}
\underbrace{\bigg[
p^\mathrm{eff}\delta_{ik}
-
2\mu^\mathrm{eff}S_{ik}
\bigg]}_{f_i}
+
\alpha
v_i
&=
0
\, .
\label{equ:momentum_balance}
\end{align}
Here, $v_i=v_i(x_i,t)$ and $\rho$ denote the fluid velocity and density, respectively. Therein, $x_i$ and $t$ represent the spatial and temporal coordinates. Furthermore, $S_{ik}=1/2(\partial v_i/\partial x_k+\partial v_k/\partial x_i)$ represents the strain-rate tensor, while $p^\mathrm{eff}=p+p^t$ and $\mu^\mathrm{eff}=\mu+\mu^t$ denote the effective pressure and viscosity, respectively, including the true fluid pressure $p$, molecular viscosity $\mu$, as well as possible turbulence-model contributions $p^t$ and $\mu^t$. Depending on the Reynolds number under consideration, the latter either vanish for laminar simulations or represent the corresponding Reynolds-averaged turbulence closure contributions. The turbulent cases are modeled using the 2003 version of Menter's Shear-Stress-Transport (SST) model, cf. \cite{menter2003ten}. The resulting aerodynamic surface force vector obtained from the pressure and viscous stresses is denoted by $f_i$.


The aerodynamic quantities of interest are the drag force $F_D$, the lift force $F_L$, and the torque $M_T$ about the cylinder axis $x_3$, cf. Fig.~\ref{fig:rotating_cylinder_2D}. Throughout this work, the corresponding optimization objectives are the minimization of drag and torque and the maximization of lift. Based on the cylinder's surface force vector $f_i$, drag and lift are obtained from the projections $F_D=-F_i\delta_{i1}$ and $F_L=F_i\delta_{i2}$, respectively, where $F_i$ denotes the integrated surface force vector acting on the cylinder boundary. The minus sign in the definition of $F_D$ accounts for the incoming flow being directed opposite to the positive $x_1$ direction. The corresponding non-dimensional coefficients are defined as
\begin{align}
    c_D = \frac{F_D}{\frac{1}{2}\rho V^2 D H} \, , \qquad \qquad
    c_L = \frac{F_L}{\frac{1}{2}\rho V^2 D H} \qquad \qquad \text{and} \qquad \qquad
    c_T = \frac{M_T}{\frac{1}{2}\rho V^2 D^2 H} \, , \label{equ:quantities_of_interest}
\end{align}
where $H$ denotes the cylinder height. The operating point is characterized by the diameter-based Reynolds number and the spinning ratio,
\begin{align}
    \mathrm{Re}_D = \frac{\rho V D}{\mu} \qquad \qquad \text{and} \qquad \qquad
    \lambda = \frac{\pi D n}{V} \, .
\end{align}
Both $\mathrm{Re}_D$ and $\lambda$ are varied systematically throughout the present study. The investigated parameter space covers Reynolds numbers ranging from $\mathrm{Re}_D=10^{1}$ to $\mathrm{Re}_D=10^{7}$ and spinning ratios between $\lambda=0$ and $\lambda=2\pi$.

For each operating point, the converged flow solution is subsequently transferred to a continuous adjoint solver. For unsteady flow simulations, all quantities are evaluated from time-averaged flow fields obtained over 100 cylinder revolutions. Sensitivity fields are computed independently for the drag, lift, and torque coefficients. Consequently, each operating point requires one primal flow simulation and three corresponding adjoint evaluations. The corresponding adjoint equations and boundary conditions follow directly from the continuous adjoint formulation and are not repeated here for brevity. Further details can be found in the authors' previous work. e.g., \cite{kuhl2019decoupling, kuhl2021adjoint_2, kuhl2025adjoint}.

Based on the primal and adjoint velocity fields, the local topological sensitivity is evaluated as
\begin{align}
    s = -\hat{v}_i v_i \, , \label{equ:sensitivity}
\end{align}
where $\hat{v}_i$ denotes the adjoint velocity field. The sensitivity represents the first-order variation of the respective objective with respect to the local Darcy-type momentum source in Eqn. \eqref{equ:momentum_balance}. The negative sign follows the conventional interpretation used in topology optimization and ensures that positive sensitivity values correspond to favourable modifications of the Darcy parameter. In a gradient-based optimization framework, the resulting field would subsequently be scaled by a suitable step size to obtain a design update. In the present work, however, no optimization loop is performed. Instead, except for dedicated validation studies, only the sign and spatial distribution of the sensitivity field are evaluated and interpreted as indicators for favourable and unfavourable locations of passive flow-control devices or additional structures in the vicinity of the rotating cylinder. Consequently, positive values indicate regions where additional momentum extraction is expected to improve the considered objective, whereas negative values indicate detrimental regions. Only for validation purposes, a dedicated forward simulation employing a sensitivity-informed source/sink distribution is carried out.

For completeness, the corresponding integral sensitivities associated with the drag, lift, and torque objectives are additionally evaluated by integrating the respective local sensitivity fields over a rectangular domain of dimensions $5D \times 5D$ centered around the cylinder and normalizing the resulting quantities analogously to the aerodynamic coefficients introduced in Eqn.~\eqref{equ:quantities_of_interest}. These integral measures provide a compact characterization of the overall sensitivity level for the considered operating point and are employed in the subsequent analysis:
\begin{align}
\hat{c}_D = \frac{S_D}{V} \, ,\qquad \qquad
\hat{c}_L = \frac{S_L}{V} \qquad \qquad \text{and} \qquad \qquad
\hat{c}_T = \frac{S_T}{V D} \, ,
\label{equ:adjoint_quantities_of_interest}
\end{align}
where $S_D$, $S_L$, and $S_T$ denote the integrals of the local sensitivity field over the prescribed $5D \times 5D$ region from Eqn.~\eqref{equ:sensitivity}, corresponding to the drag, lift, and torque objectives, respectively.

All primal and adjoint flow simulations are performed using the finite-volume software package FreSCo$^+$. The solver employs a collocated variable arrangement on unstructured grids and discretizes the governing equations by means of a second-order accurate finite-volume method. Pressure--velocity coupling is achieved through a SIMPLE-type algorithm. Further details regarding the numerical framework can be found in \cite{rung2009challenges, stuck2012adjoint, manzke2018development, schubert2019analysis}

The flow is assumed to be steady for $\mathrm{Re}_D=10$ and unsteady for all higher Reynolds numbers. Simulations are performed in laminar mode for $\mathrm{Re}_D<10^5$ and in RANS mode for $\mathrm{Re}_D\geq10^5$. Throughout the adjoint computations, the frozen-turbulence assumption is adopted, i.e., the turbulence quantities are not included in the adjoint formulation, cf. \cite{soto2004adjoint, othmer2008continuous, stuck2013adjoint}. No wall functions are employed in primal and adjoint mode; instead, all simulations are performed in a low-Reynolds-number formulation with the near-wall boundary layer fully resolved, cf. \cite{kuhl2024continuous}. Time-accurate forward simulations employ an adaptive time-stepping strategy that maintains a target Courant number of approximately five in a least-squares sense throughout the computational domain.

Velocity boundary conditions with constant velocity values are prescribed at the inlet as well as at the upper and lower far-field boundaries, while a homogeneous fixed-pressure boundary condition is imposed at the outlet. Symmetry boundary conditions are applied in the out-of-plane direction, yielding a two-dimensional flow configuration. For the RANS simulations, a turbulence intensity of $1\,\%$ and a normalized eddy viscosity of unity are prescribed at the velocity boundaries. All simulations are initialized by uniformly transferring the prescribed boundary values into the computational domain.

A formal numerical uncertainty assessment is performed for each operating point following the procedure proposed by \cite{ecca2014procedure}. Therein, the observed grid convergence is approximated by a two-term error expansion $\phi(h)=\phi_0 + a h^p + b h^{p+1}$, where $\phi_0$ denotes the extrapolated grid-independent solution of the quantities of interest, cf. Eqn.~\eqref{equ:quantities_of_interest}, $h$ denotes a characteristic mesh size, and $p$ is the observed order of convergence.

The corresponding discretization uncertainties are estimated from a sequence of six systematically refined grids. A dedicated grid family is generated for each Reynolds number under consideration, whereas all spinning ratios associated with a given Reynolds number are evaluated on the same meshes. The near-wall resolution is selected such that $y^+<1$ is maintained even for the highest spinning ratios considered in the study. Consequently, only the Reynolds number affects the grid generation process. Since the grid spacing remains constant along the cylinder circumference, however, local $y^+$ values may differ noticeably between the upper and lower cylinder sides due to the varying relative velocity resulting from the superposition of the free-stream and circumferential velocities.
In contrast, the far-field discretization remains unchanged throughout the study and is based on a generously sized computational domain surrounding the cylinder. To minimize blockage effects and boundary-condition influences, all outer boundaries are positioned 100 cylinder diameters away from the cylinder center, resulting in a square computational domain with an edge length of 200 diameters. The corresponding refinement region extends well beyond the immediate cylinder vicinity and wake in order to adequately resolve both the primal flow field and the upstream information transport associated with the adjoint equations. This ensures a consistent representation of the resulting sensitivity fields across the entire operating envelope.

\begin{figure}[!ht]
\centering
\subfigure[]{\includegraphics[width=0.22\textwidth, trim=11cm 0cm 11cm 0cm, clip]{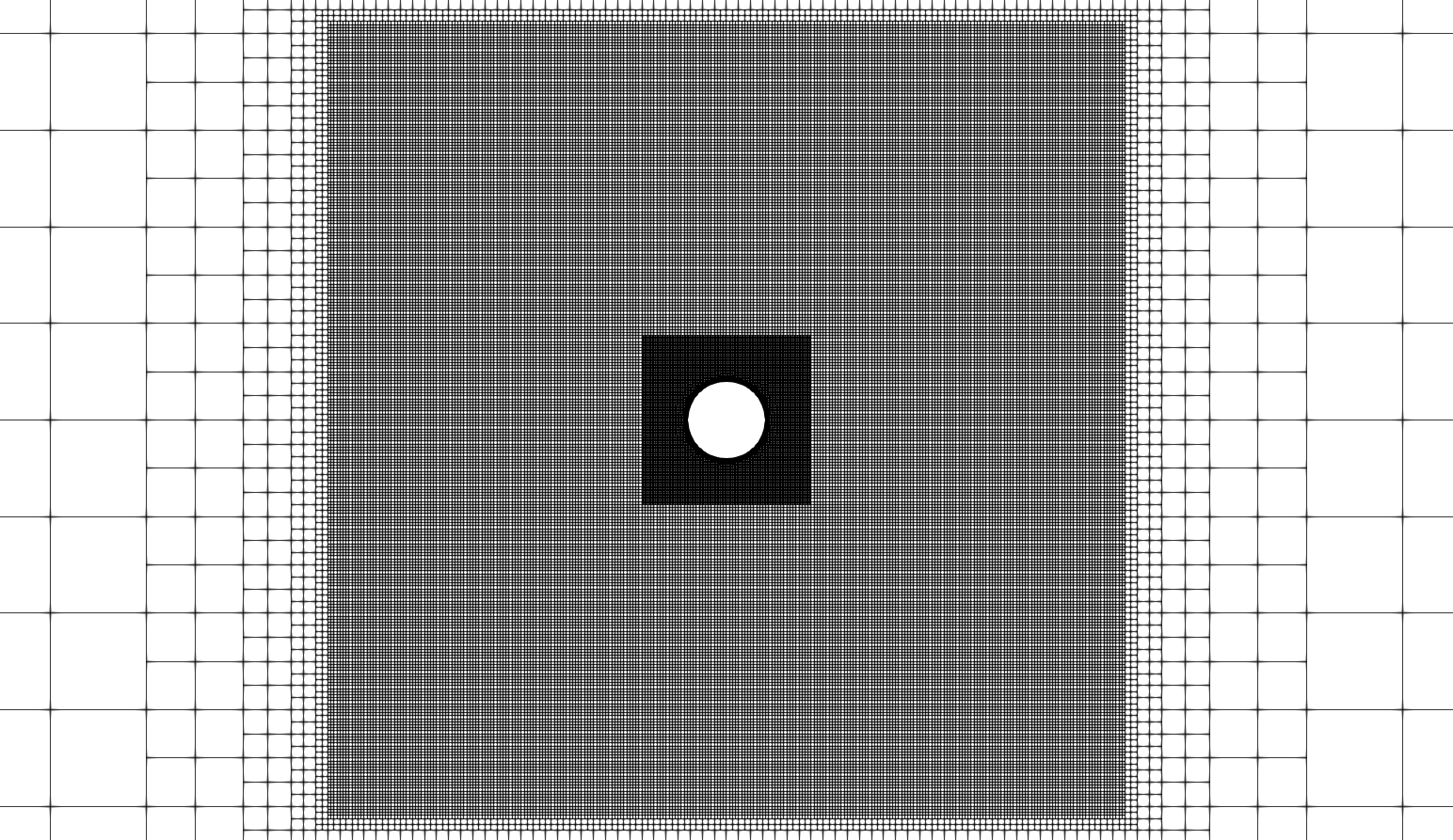}}
\subfigure[]{\includegraphics[width=0.22\textwidth, trim=11cm 0cm 11cm 0cm, clip]{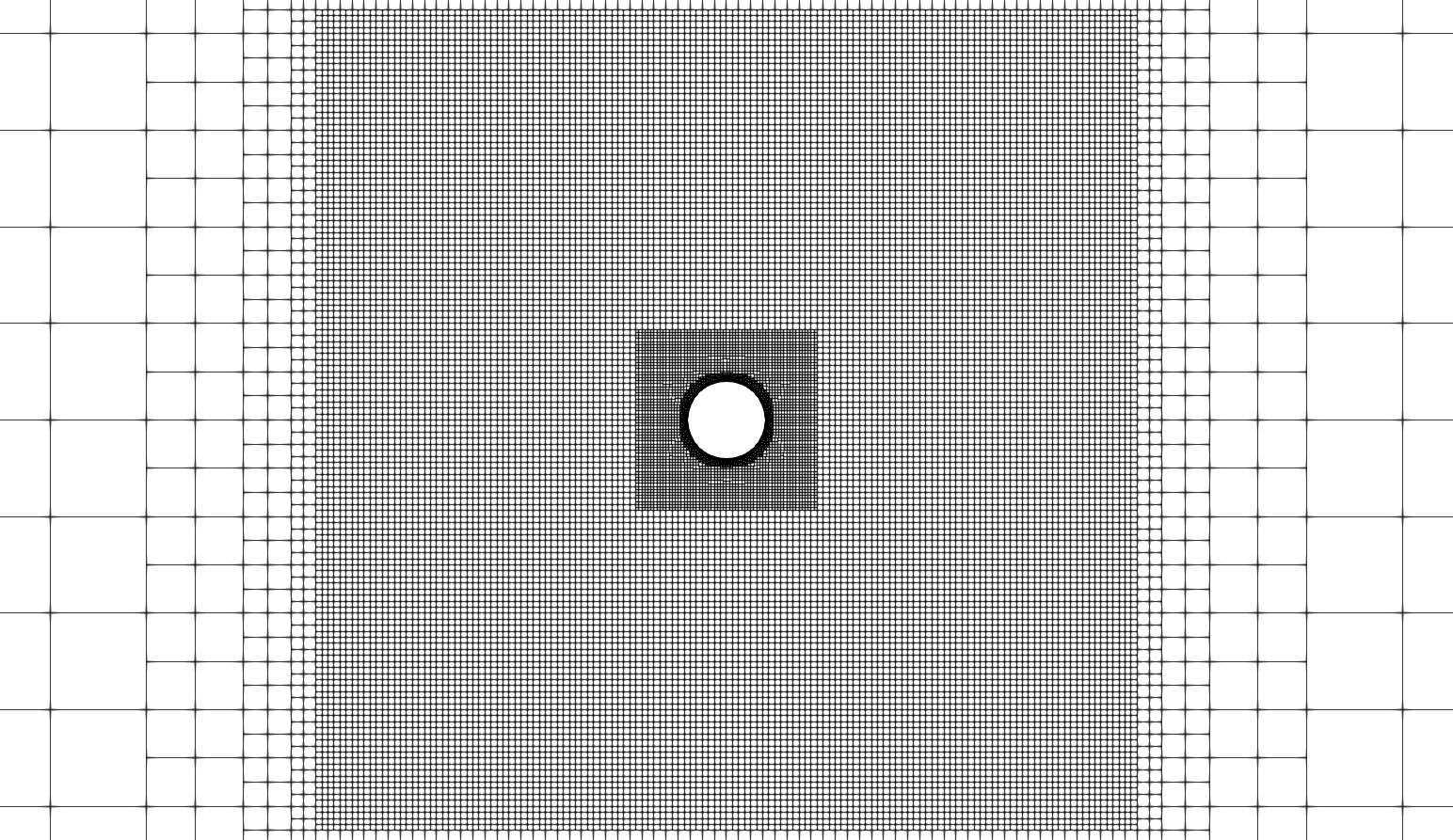}}
\subfigure[]{\includegraphics[width=0.22\textwidth, trim=11cm 0cm 11cm 0cm, clip]{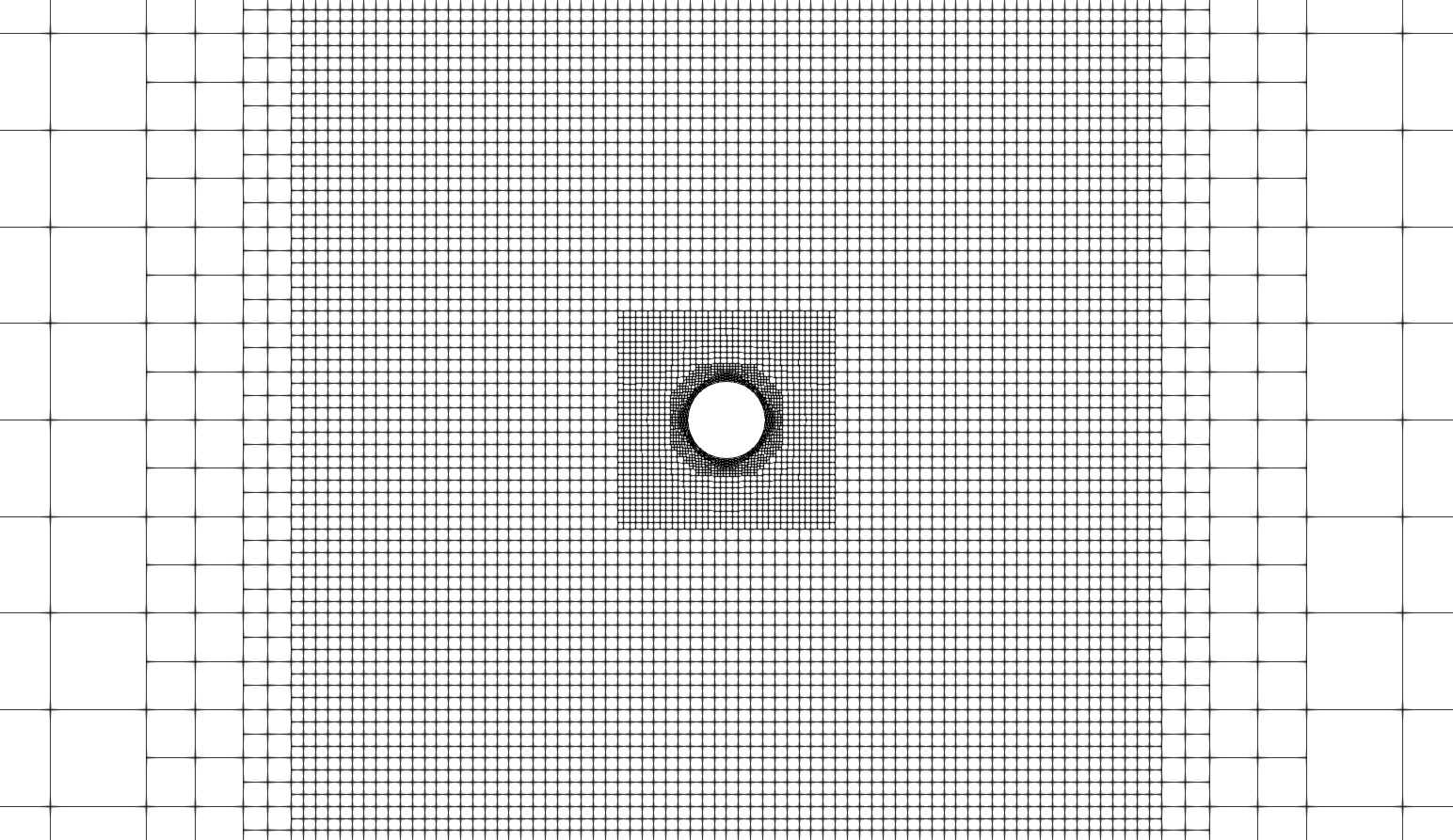}}
\subfigure[]{\includegraphics[width=0.22\textwidth, trim=11cm 0cm 11cm 0cm, clip]{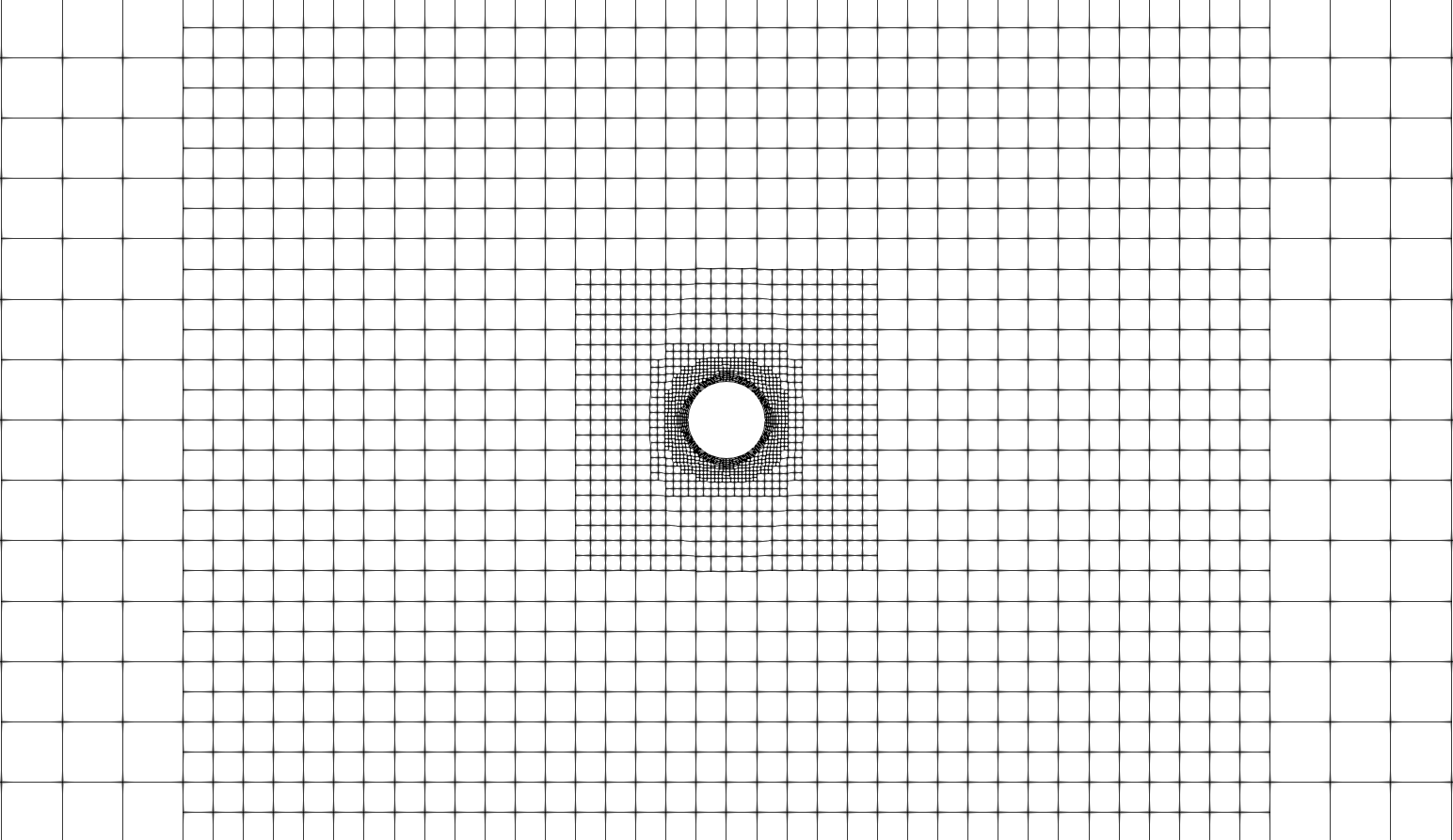}}
\caption{Near-wall discretization in the vicinity of the cylinder for four representative members of the six-level refinement sequence at $\mathrm{Re}_D=10$. Panels (a)--(d) show refinement levels 1--4, respectively. The coarsest (level 5) and finest (level 0) grids are omitted for clarity.}
\label{fig:grid_refinement}
\end{figure}

Figure~\ref{fig:grid_refinement} illustrates four representative members of the six-grid refinement sequence for the case $\mathrm{Re}_D=10$. The coarsest and finest grids are omitted. All flow fields and sensitivity distributions presented in this paper are evaluated on the finest grid of the corresponding refinement sequence. This grid is one additional refinement level finer than the finest grid displayed in Fig.~\ref{fig:grid_refinement}(a). The figure is restricted to the cylinder vicinity and therefore only shows a small portion of the computational domain.

In total, this results in $7 \times 5 \times 6 = 210$ primal simulations for the uncertainty assessment, covering seven Reynolds numbers, five spinning ratios, and six grid levels. Including the three adjoint problems associated with drag, lift, and torque, this corresponds to a total of $840$ primal and adjoint simulations for the uncertainty assessment.

\section{Validation}
\label{sec:validation}

Before investigating the complete operating envelope, the predictive capability of the proposed sensitivity-based methodology is assessed using two representative validation studies. The first validation remains fully consistent with the underlying topological-sensitivity concept and applies a sensitivity-informed Darcy-type source/sink distribution in a subsequent forward simulation. The second validation considers an actual geometric modification in the form of a passive flow-control device derived from the computed sensitivity field. Together, both studies evaluate whether the predicted favourable and unfavourable regions correspond to measurable changes in the aerodynamic performance of the rotating cylinder.

\subsection{Source/Sink Validation}
\label{subsec:source_sink_validation}

A representative operating point at $\mathrm{Re}_D = 10$ and $\lambda = \pi$ is selected for the initial validation study. This operating point exhibits a steady laminar flow solution and therefore provides a suitable basis for a first assessment of the proposed methodology. The corresponding primal and adjoint solutions are first evaluated according to the procedure described in Section~\ref{sec:numerical_method}. As discussed previously, the Darcy parameter vanishes throughout all primal and adjoint calculations, i.e., $\alpha=0$ in the entire computational domain.

\begin{figure}[!ht]
    \centering
    \iftoggle{tikzExternal}{
    \input{./tikz/exemplary_flow_field.tikz}}{
    \includegraphics{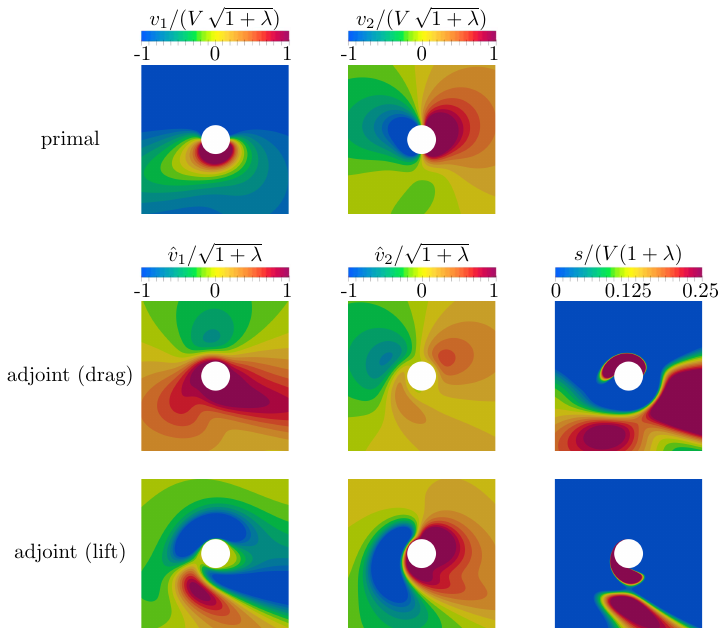}}
    \caption{Representative normalized primal, adjoint, and non-negative sensitivity fields for the validation case at $\mathrm{Re}_D=10$ and $\lambda=\pi$.}
    \label{fig:exemplariy_flow_field}
\end{figure}

Figure~\ref{fig:exemplariy_flow_field} provides an impression of the resulting primal, adjoint, and sensitivity fields for the validation case. Since the primal solution is independent of the selected objective function, the corresponding velocity components are shown only once in the first row. The second and third rows present the adjoint velocity components and the resulting non-negative sensitivity fields associated with the drag and lift objectives, respectively. As expected, the different objective functions lead to distinctly different adjoint and sensitivity distributions despite being based on the same underlying flow field.

Subsequently, the computed sensitivity fields are converted into Darcy-type source/sink distributions by prescribing the Darcy parameter according to $\alpha=\beta s$, where $s$ denotes the sensitivity field from Eqn.~\eqref{equ:sensitivity}, exemplarily illustrated for the drag and lift objectives in Fig.~\ref{fig:exemplariy_flow_field}, and $\beta=10^{-3}$ is a prescribed step-size parameter of dimension $[\beta] = N^2/([J] V D^2 H)$, where $[J]$ denotes the dimension of the cost functional, e.g., $[J] = N$ for the drag and lift objectives. In contrast to typical topology-optimization procedures, the resulting Darcy distributions are applied throughout the entire computational domain rather than being restricted to selected regions. Furthermore, both positive and negative sensitivity values are retained. Consequently, the resulting Darcy distributions contain momentum sinks as well as momentum sources. This procedure is adopted exclusively for validation purposes in order to assess the predictive capability of the sensitivity field throughout the entire computational domain.

Three additional forward simulations are performed using Darcy distributions derived from the drag, lift, and torque sensitivities, respectively. The resulting changes in the aerodynamic coefficients from Eqn.~\eqref{equ:quantities_of_interest} are evaluated relative to the unmodified reference solution. In addition to the individual coefficients $c_D$, $c_L$, and $c_T$, the lift-to-drag ratio $c_L/c_D$ is considered as an integral measure of aerodynamic efficiency. Since each Darcy distribution directly follows the sensitivity field of one specific objective, the corresponding diagonal entries in Tab.~\ref{tab:source_sink_validation} provide the primary validation of the computed sensitivity directions. The remaining entries quantify cross-effects between drag, lift, torque, and aerodynamic efficiency.
\begin{table}[!ht]
\centering
\caption{Relative changes in the aerodynamic coefficients and lift-to-drag ratio resulting from sensitivity-informed Darcy source/sink distributions for the validation case at $\mathrm{Re}_D=10$ and $\lambda=\pi$.}
\label{tab:source_sink_validation}
\begin{tabular}{lcccc}
\toprule
Darcy distribution based on &
$\Delta c_D/c_D$ [\%] &
$\Delta c_L/c_L$ [\%] &
$\Delta c_T/c_T$ [\%] &
$\Delta(c_L/c_D)/(c_L/c_D)$ [\%] \\
\midrule
Drag sensitivity   &  -5.145 &  -0.728 &  -0.043 &  +4.656 \\
Lift sensitivity   &  +6.488 &  +1.823 &  +0.054 &  -4.381 \\
Torque sensitivity &  -0.025 &  -0.031 &  -0.012 &  -0.006 \\
\bottomrule
\end{tabular}
\end{table}
The results in Tab.~\ref{tab:source_sink_validation} confirm the predictive capability of the computed sensitivity fields. For all three objective functions, the diagonal entries exhibit the expected sign and therefore indicate that the sensitivity formulation correctly predicts the response of the respective aerodynamic quantity. However, the corresponding Darcy distributions contain both momentum sinks and momentum sources. While this approach provides a rigorous mathematical validation of the sensitivity formulation, the resulting perturbations do not directly correspond to practical passive-flow-control devices or additional structures.

To address this aspect, a second validation study is performed in which only positive sensitivity values are retained. Accordingly, the Darcy parameter is prescribed according to $\alpha=\beta \max(s,0)$, such that only momentum sinks are introduced while all momentum sources are suppressed. Furthermore, the Darcy distribution is deactivated within a distance of $D/10$ from the cylinder surface in order to avoid direct modifications of the boundary-layer region. The resulting perturbation is therefore more closely related to practical passive-flow-control devices, porous inserts, or additional structures. Most importantly, the resulting validation no longer relies on artificial momentum injection and therefore provides a more physically interpretable assessment of the sensitivity fields. The corresponding results are summarized in Tab.~\ref{tab:sink_only_validation}.
\begin{table}[!ht]
\centering
\caption{Relative changes in the aerodynamic coefficients and lift-to-drag ratio resulting from sink-only Darcy distributions for the validation case at $\mathrm{Re}_D=10$ and $\lambda=\pi$.}
\label{tab:sink_only_validation}
\begin{tabular}{lcccc}
\toprule
Darcy distribution based on &
$\Delta c_D/c_D$ [\%] &
$\Delta c_L/c_L$ [\%] &
$\Delta c_T/c_T$ [\%] &
$\Delta(c_L/c_D)/(c_L/c_D)$ [\%] \\
\midrule
Drag sensitivity   &  -4.016 &  -0.683 &  -0.051 &  +3.472 \\
Lift sensitivity   &  -0.341 &  +0.108 &  +0.006 &  +0.450 \\
Torque sensitivity &  -0.202 &  -0.044 &  -0.004 &  +0.158 \\
\bottomrule
\end{tabular}
\end{table}
Both validation strategies consistently confirm the predictive capability of the computed sensitivity fields. For the signed source/sink distributions, the strongest effect is observed for the drag objective, where the drag coefficient decreases by approximately five percent. Likewise, the lift-based perturbation increases the lift coefficient, whereas the torque-based perturbation reduces the torque coefficient. The corresponding off-diagonal entries reveal the inherent coupling between the aerodynamic objectives. In particular, the drag- and lift-based perturbations exhibit an antagonistic behaviour, where improvements in one quantity are accompanied by a deterioration of the other. This behaviour is particularly evident in the lift-to-drag ratio, which increases for the drag-based perturbation but decreases for the lift-based perturbation.

The sink-only validation leads to qualitatively similar trends, although the resulting changes are generally smaller in magnitude. This behaviour is expected since only a subset of the sensitivity information is retained. Most importantly, the physically more relevant sink-only perturbations preserve the predicted directions of the objective changes even though all momentum sources are omitted. The drag-based perturbation still reduces the drag coefficient by approximately four percent while simultaneously increasing the lift-to-drag ratio by more than three percent. Furthermore, the lift-based perturbation leads to a slight increase in lift while simultaneously reducing drag, resulting in a moderate improvement of the lift-to-drag ratio. For the considered operating point, the torque-based perturbations lead only to minor changes in all aerodynamic coefficients.

Overall, both validation approaches consistently confirm the predicted sensitivity directions and thereby provide confidence in the subsequent interpretation of the sensitivity fields across the investigated operating envelope. Moreover, the agreement between the signed and sink-only perturbations indicates that the computed sensitivity fields remain meaningful when interpreted in the context of practical passive-flow-control devices and additional structures.

\subsection{Structural Validation}
\label{subsec:structural_validation}
While the previous validation studies assessed the predictive capability of the sensitivity fields through distributed Darcy-type source and sink terms, a second validation is performed using an actual geometric modification of the computational domain. The objective of this study is to investigate whether the computed sensitivity fields can also provide meaningful guidance for the placement of passive flow-control devices or additional structures.

\begin{figure}[!ht]
    \centering
    \iftoggle{tikzExternal}{
    \input{./tikz/combined_sensitivity_field.tikz}}{
    \includegraphics{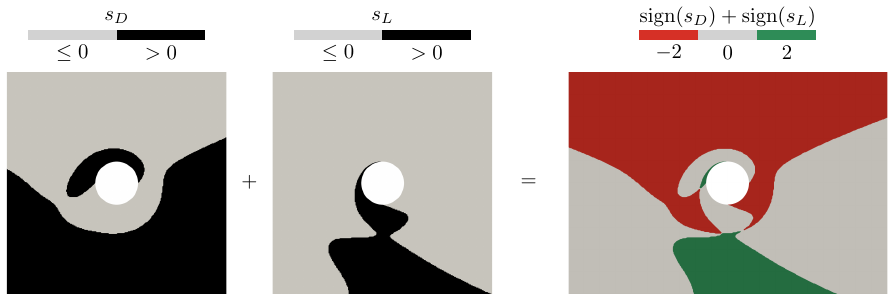}}
    \caption{Sign of the drag sensitivity, sign of the lift sensitivity, and resulting combined sign field for $\mathrm{Re}_D=10$ and $\lambda=\pi$.}
    \label{fig:combined_sensitivity_fields}
\end{figure}

To this end, the drag and lift sensitivities are considered simultaneously for the representative operating point at $\mathrm{Re}_D=10$ and $\lambda=\pi$. Since the objective of the present validation is to derive a passive flow-control structure, only the sign information of the respective sensitivity fields is retained. Figure~\ref{fig:combined_sensitivity_fields} therefore illustrates the sign of the drag sensitivity, the sign of the lift sensitivity, and the resulting combined sign field obtained by superimposing both distributions. The latter identifies regions where the sensitivity information simultaneously suggests drag reduction and lift enhancement and therefore serves as the basis for the subsequent design proposal.
For this purpose, the interface separating favourable and unfavourable regions is extracted and converted into a closed contour. Since the objective is to derive a passive structure interacting with the cylinder wake, only the contour segment located directly downstream of the cylinder is retained, whereas the more remote regions identified by the combined sign field are discarded. To avoid direct interference with the immediate cylinder vicinity, a circular exclusion zone with a diameter of $1.1D$ is introduced around the cylinder and subtracted from the extracted contour. The resulting geometry is subsequently smoothed and transformed into a streamlined structure using an external CAD environment.

\begin{figure}[!ht]
    \centering
    \iftoggle{tikzExternal}{
    \input{./tikz/structural_validation_geometry.tikz}}{
    \includegraphics{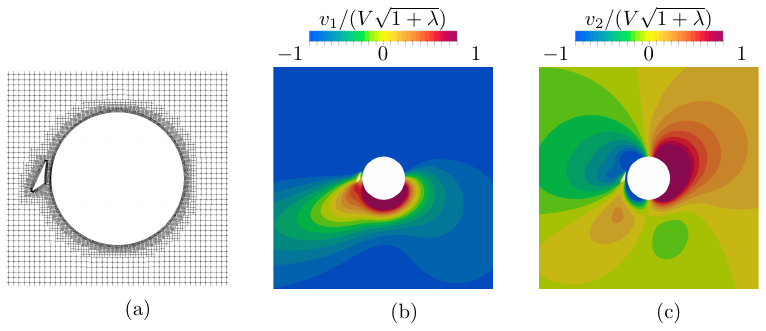}}
    \caption{Structural validation case at $\mathrm{Re}_D=10$ and $\lambda=\pi$: (a) body-fitted near-wall mesh around the sensitivity-informed structure, (b) streamwise velocity component $v_1$, and (c) transverse velocity component $v_2$ obtained from the corresponding forward simulation.}
    \label{fig:structural_validation_geometry}
\end{figure}

The resulting structure is incorporated into the computational domain and discretized using a newly generated body-fitted mesh. Figure~\ref{fig:structural_validation_geometry} illustrates the near-wall mesh around the modified configuration together with the corresponding velocity components obtained from the subsequent forward simulation.

The resulting aerodynamic coefficients are subsequently compared to the corresponding reference solution. Since the additional structure contributes to the overall force and moment balance, two different evaluations are considered. First, only the loads acting on the original cylinder surface are taken into account. Second, the contributions of both the cylinder and the sensitivity-informed structure are included in the force and moment balance. The corresponding results are summarized in Tab.~\ref{tab:structural_validation}.
\begin{table}[!ht]
\centering
\caption{Relative changes in the aerodynamic coefficients and lift-to-drag ratio resulting from the sensitivity-informed structure at $\mathrm{Re}_D=10$ and $\lambda=\pi$.}
\label{tab:structural_validation}
\begin{tabular}{lcccc}
\toprule
Configuration &
$\Delta c_D/c_D$ [\%] &
$\Delta c_L/c_L$ [\%] &
$\Delta c_T/c_T$ [\%] &
$\Delta(c_L/c_D)/(c_L/c_D)$ [\%] \\
\midrule
Cylinder only            & -66.531 &  +2.840 & +64.528 & +207.267 \\
Cylinder + add-on        & -61.858 & -50.512 & -13.108 & +29.748 \\
\bottomrule
\end{tabular}
\end{table}

The cylinder-only evaluation is the relevant quantity for validating the sensitivity prediction, since the adjoint objective is defined on the original cylinder surface. Accordingly, the sensitivity-informed structure produces the expected qualitative response: the cylinder drag is strongly reduced while the cylinder lift is increased. When the loads acting on the additional structure are included, the total force balance changes substantially. This is expected, since the aerodynamic loads of the newly introduced structure are not part of the original objective functional and therefore cannot be anticipated by the sensitivity field. Nevertheless, even the combined cylinder--structure evaluation still shows a considerable improvement of the lift-to-drag ratio.

Together with the source/sink validation, this geometric validation indicates that the computed sensitivity fields can be used both as local first-order indicators and as qualitative guidance for passive flow-control design. The transferability of this design strategy to a practically relevant high-Reynolds-number ($\mathrm{Re}_D=10^6$) operating point is investigated in the later Sec. \ref{subsec:operating_point_transferability}.

\subsection{Integral Aerodynamic Quantities}
\label{subsec:integral_aerodynamic_quantities}
\begin{figure}[!ht]
    \centering
    \includegraphics[scale=0.7]{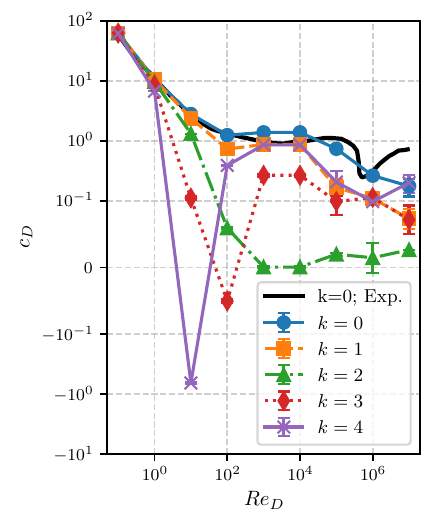}
    \includegraphics[scale=0.7]{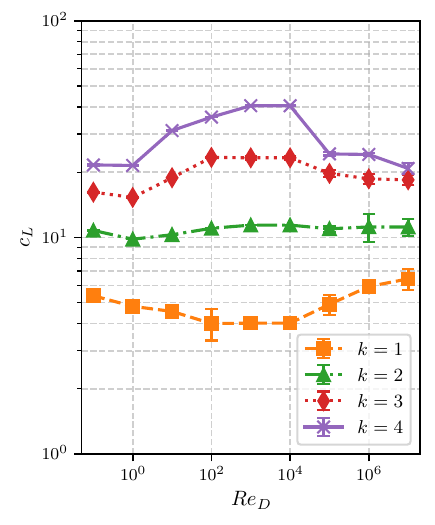}
    \includegraphics[scale=0.7]{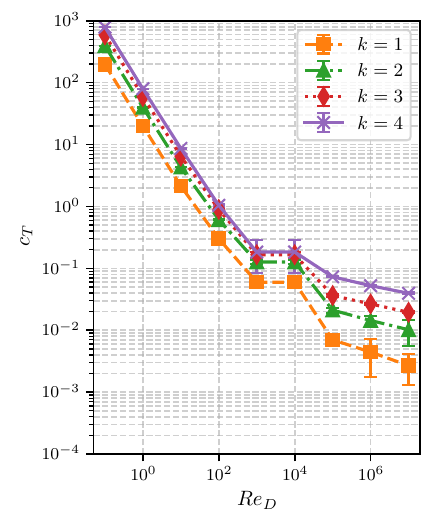}
    \caption{Integral aerodynamic quantities as functions of Reynolds number and spinning ratio: (a) drag coefficient, (b) lift coefficient, and (c) torque coefficient. Error bars indicate the estimated discretization uncertainty. Experimental reference data are included for the stationary cylinder.}
    \label{fig:integral_quantities}
\end{figure}

Before discussing the resulting sensitivity distributions, the integral aerodynamic response of the rotating cylinder is briefly reviewed. Figure~\ref{fig:integral_quantities} summarizes the drag, lift, and torque coefficients obtained throughout the investigated operating envelope. For the stationary cylinder ($\lambda=0$), available experimental reference data are additionally included. Furthermore, the error bars indicate the estimated discretization uncertainties obtained from the grid-convergence studies described in Section~\ref{sec:numerical_method}.
Overall, the numerical results reproduce the expected trends of the rotating cylinder over a wide range of Reynolds numbers and spinning ratios. The torque coefficient decreases systematically with increasing Reynolds number and approaches zero in the turbulent regime. In contrast, the lift coefficient increases strongly with increasing spinning ratio and exhibits only a comparatively weak Reynolds-number dependence. The drag coefficient generally decreases with increasing Reynolds number, reflecting the well-known evolution towards the drag-crisis regime. For sufficiently large spinning ratios, slightly negative drag coefficients are obtained in the laminar flow regime, indicating a net propulsive force generated by the rotating cylinder. Similar behaviour has previously been reported in numerical investigations by \cite{stojkovic2002effect}.

Good agreement with the available experimental reference data is obtained for the stationary cylinder, particularly at low and moderate Reynolds numbers. For Reynolds numbers exceeding approximately $\mathrm{Re}_D=10^5$, both the estimated discretization uncertainty and the deviation from the experimental data increase noticeably. This behaviour reflects the increasing complexity of the underlying flow physics and the corresponding sensitivity to modelling and discretization assumptions. Nevertheless, the numerical uncertainties remain sufficiently small compared to the overall parameter-induced variations and therefore permit a meaningful interpretation of both the aerodynamic response and the corresponding sensitivity distributions.

\begin{figure}[!ht]
    \centering
    \includegraphics[scale=0.7]{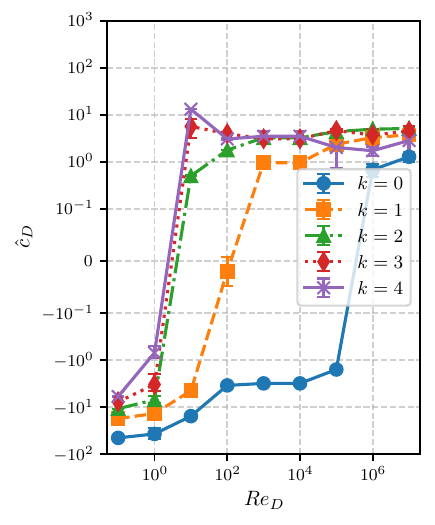}
    \includegraphics[scale=0.7]{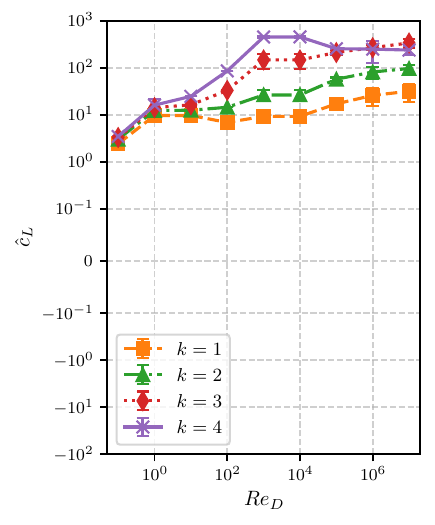}
    \includegraphics[scale=0.7]{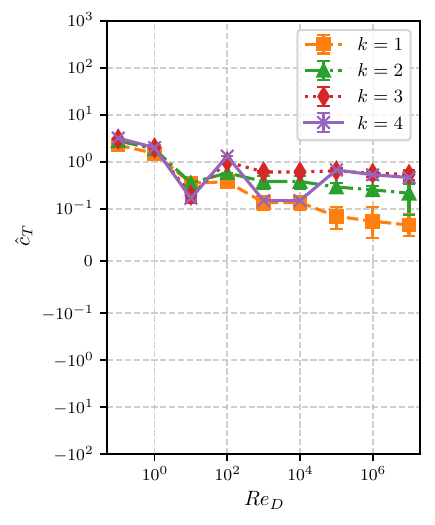}
    \caption{Integral adjoint quantities associated with the drag, lift, and torque objectives as functions of Reynolds number and spinning ratio. The local sensitivity fields are integrated over a $5D \times 5D$ region centered around the cylinder. Error bars indicate the estimated spatial discretization uncertainty.}
    \label{fig:integral_quantities_adjoint}
\end{figure}

In addition to the aerodynamic coefficients, Fig.~\ref{fig:integral_quantities_adjoint} summarizes the corresponding integral objective values of the adjoint problems, cf. Eqn.~\eqref{equ:adjoint_quantities_of_interest}. Since the adjoint equations are solved independently for drag, lift, and torque, the resulting quantities provide a compact characterization of the overall sensitivity response throughout the investigated operating envelope. To facilitate a direct comparison of the overall sensitivity levels associated with the different objectives, all three panels are presented using an identical ordinate scale.
The integral adjoint quantities exhibit a pronounced dependence on both Reynolds number and spinning ratio. For the stationary cylinder, the drag-related integral sensitivity is predominantly negative at low Reynolds numbers and changes sign within the transitional Reynolds-number range. In contrast, for the rotating cylinders, the sign change occurs already near the transition from steady to unsteady flow. At higher Reynolds numbers, the drag-related integral sensitivities associated with the rotating cylinders approach an approximately constant positive level, whereas the stationary-cylinder response remains distinctly lower.

The lift-related integral sensitivity generally increases with Reynolds number and spinning ratio. The strongest increase occurs for the larger spinning ratios, which also yield the highest sensitivity levels in the high-Reynolds-number regime. In contrast, the torque-related integral sensitivity decreases markedly between the lowest Reynolds numbers and approximately $\mathrm{Re}_D=10$. Beyond this range, it remains at a comparatively low level and exhibits only moderate variations with Reynolds number and spinning ratio.
Overall, the lift objective produces the largest integral sensitivity throughout most of the investigated operating envelope, while the torque objective generally exhibits the smallest values. The drag-related sensitivity assumes intermediate values at higher Reynolds numbers, although its sign change at lower Reynolds numbers reflects a more complex dependence on the operating condition. The common ordinate scale employed in Fig.~\ref{fig:integral_quantities_adjoint} has therefore been chosen deliberately to facilitate this direct comparison.
This qualitative ordering is consistent with the source/sink validation study presented in Section~\ref{subsec:source_sink_validation}. There, the unmodified sensitivity fields from Eqn.~\eqref{equ:sensitivity} produced the largest response for the lift objective and the smallest response for the torque objective, while the drag response assumed an intermediate level, cf. Tab.~\ref{tab:source_sink_validation}. It should further be emphasized that the integral adjoint quantities shown in Fig.~\ref{fig:integral_quantities_adjoint} are obtained by integrating the original sensitivity fields over the prescribed $5D \times 5D$ region surrounding the cylinder. Consequently, no clipping of negative sensitivities or exclusion of the near-cylinder region, as employed in the dedicated validation studies, is applied.

The following section investigates the spatial structure of these sensitivities and its evolution throughout the investigated Reynolds-number and spinning-ratio ranges.

\section{Sensitivity Spectra}
\label{sec:sensitivity_spectra}
Having established the validity of the proposed sensitivity formulation and characterized the corresponding aerodynamic response, this section investigates the resulting sensitivity distributions throughout the complete operating envelope. Particular emphasis is placed on the influence of Reynolds number and spinning ratio on the spatial structure of the sensitivity fields.

For each operating point, the magnitude of the primal velocity field is considered together with the corresponding sensitivity distributions for drag, lift, and torque. For unsteady flow cases, time-averaged velocity fields are evaluated. In addition, the combined drag--lift sign field introduced previously in Fig.~\ref{fig:combined_sensitivity_fields} is considered in order to identify regions that simultaneously favour drag reduction and lift enhancement.

\begin{figure}[!ht]
    \centering
    \iftoggle{tikzExternal}{
    \input{tikz/velocity_spectrum.tikz}}{
    \includegraphics{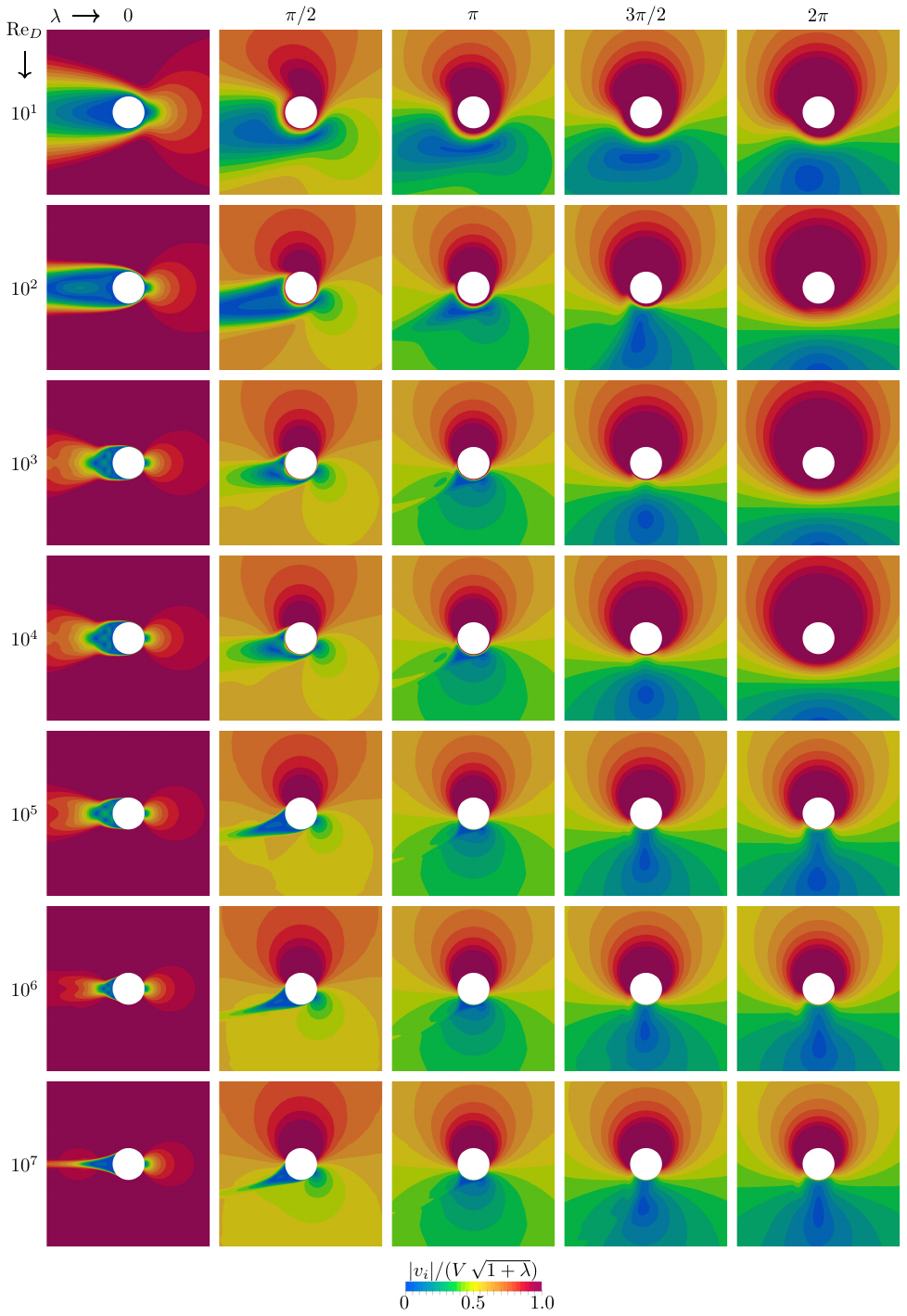}
    }
    \caption{Magnitude of the primal velocity field for all investigated Reynolds numbers and spinning ratios.}
    \label{fig:velocity_spectrum}
\end{figure}

Figure~\ref{fig:velocity_spectrum} illustrates the normalized velocity magnitude. With increasing Reynolds number, the wake becomes progressively narrower and the strongest velocity gradients are increasingly confined to the vicinity of the cylinder surface. At the same time, increasing cylinder rotation gradually replaces the nearly symmetric wake by a pronounced downward-deflected jet. In particular, for $\lambda\geq\pi$ and $\mathrm{Re}_D\geq10^4$, the normalized velocity fields exhibit remarkably similar large-scale structures, indicating that the mean flow topology becomes nearly independent of the Reynolds number while remaining influenced by the spinning ratio.

\begin{figure}[!ht]
    \centering
    \iftoggle{tikzExternal}{
    \input{tikz/drag_sensitivity_spectrum.tikz}}{
    \includegraphics{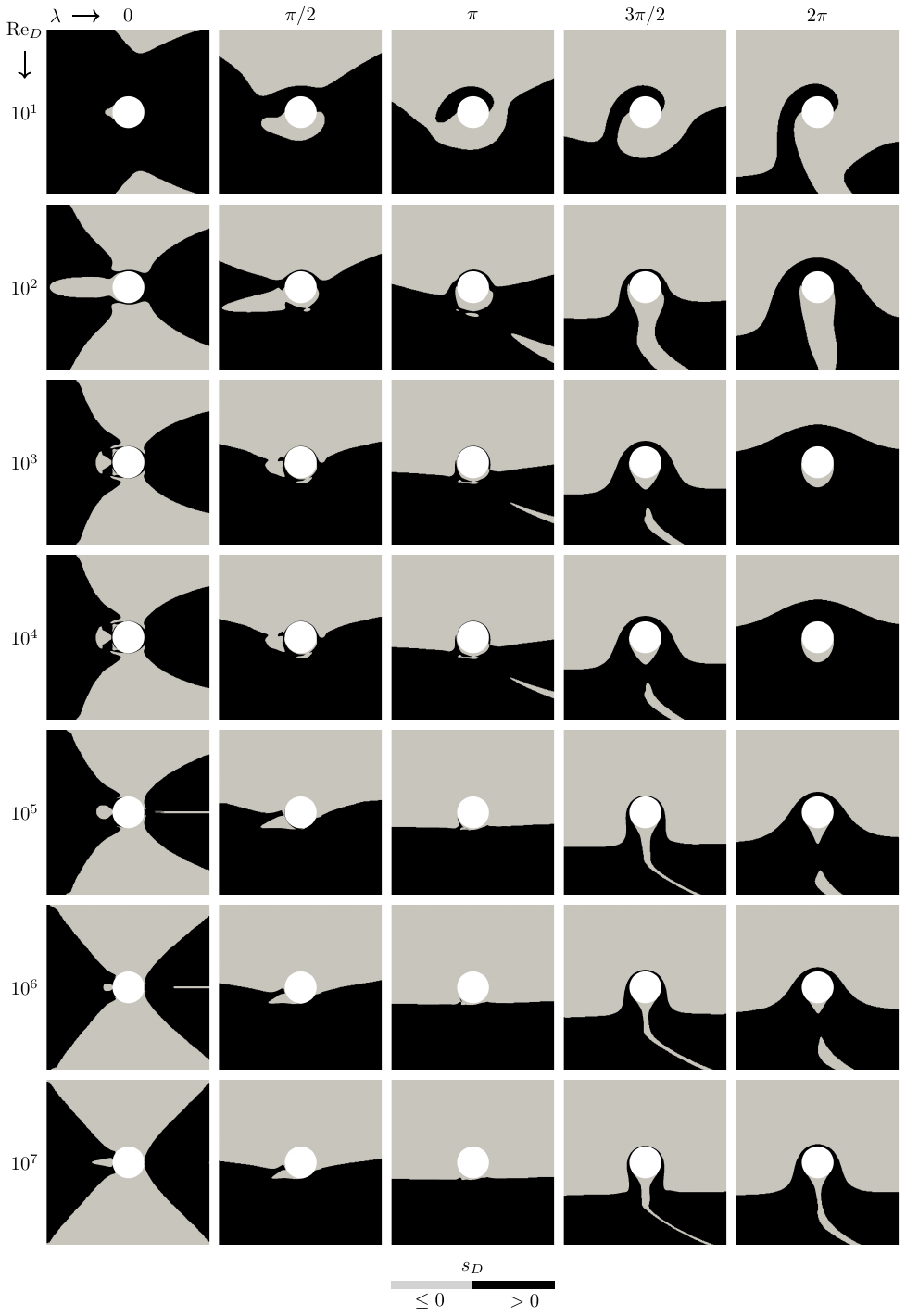}
    }
    \caption{Sign of the drag sensitivity for all investigated Reynolds numbers and spinning ratios. Black regions indicate favourable locations, whereas grey regions indicate unfavourable locations.}
    \label{fig:drag_sensitivity_spectrum}
\end{figure}

Figure~\ref{fig:drag_sensitivity_spectrum} summarizes the corresponding drag sensitivity fields. For the stationary cylinder, the sensitivity distribution is symmetric with respect to the horizontal cylinder axis, whereas increasing cylinder rotation progressively breaks this symmetry. Overall, the drag sensitivity exhibits only a comparatively weak dependence on the Reynolds number, while remaining strongly governed by the spinning ratio. In particular, the sensitivity fields for $\lambda=\pi/2$ and $\lambda=\pi$ are characterized by a distinct favourable region below the cylinder. For $\lambda=3\pi/2$ and $\lambda=2\pi$, this favourable region extends into a curtain-like structure surrounding the cylinder, while the unfavourable sensitivity is largely confined to the downstream jet.

\begin{figure}[!ht]
    \centering
    \iftoggle{tikzExternal}{
    \input{tikz/lift_sensitivity_spectrum.tikz}}{
    \includegraphics{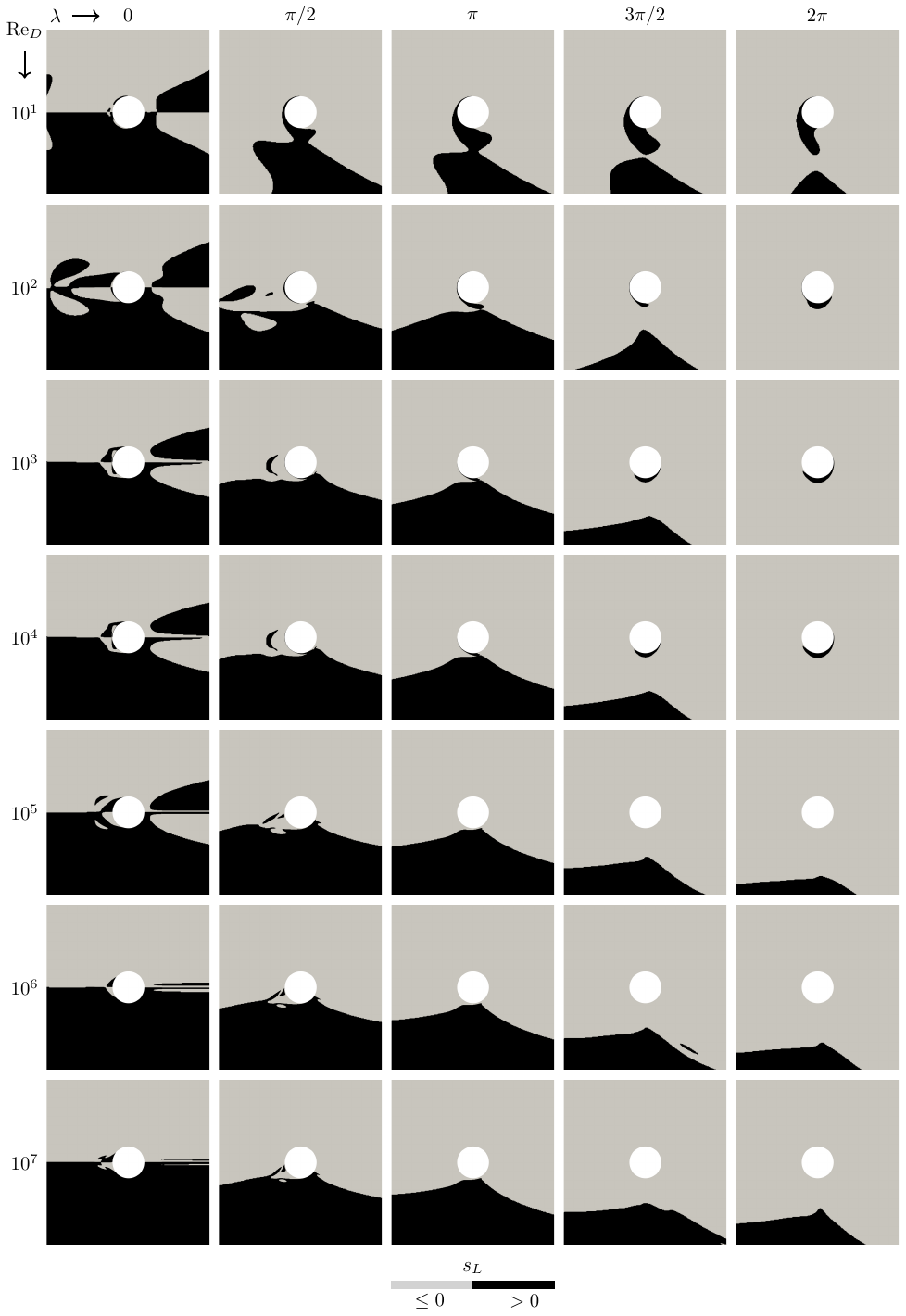}
    }
    \caption{Sign of the lift sensitivity for all investigated Reynolds numbers and spinning ratios. Black regions indicate favourable locations, whereas grey regions indicate unfavourable locations.}
    \label{fig:lift_sensitivity_spectrum}
\end{figure}

Figure~\ref{fig:lift_sensitivity_spectrum} presents the corresponding lift sensitivity fields. In contrast to the drag sensitivity, the lift sensitivity is already asymmetric for the stationary cylinder, reflecting the directional nature of the lift objective. Similar to the drag sensitivity, the large-scale topology exhibits only a comparatively weak dependence on the Reynolds number. For rotating cylinders, the favourable sensitivity is consistently located below the cylinder. With increasing spinning ratio, this favourable region becomes progressively smaller, while the remaining flow field is predominantly characterized by unfavourable sensitivity.

\begin{figure}[!ht]
    \centering
    \iftoggle{tikzExternal}{
    \input{tikz/torque_sensitivity_spectrum.tikz}}{
    \includegraphics{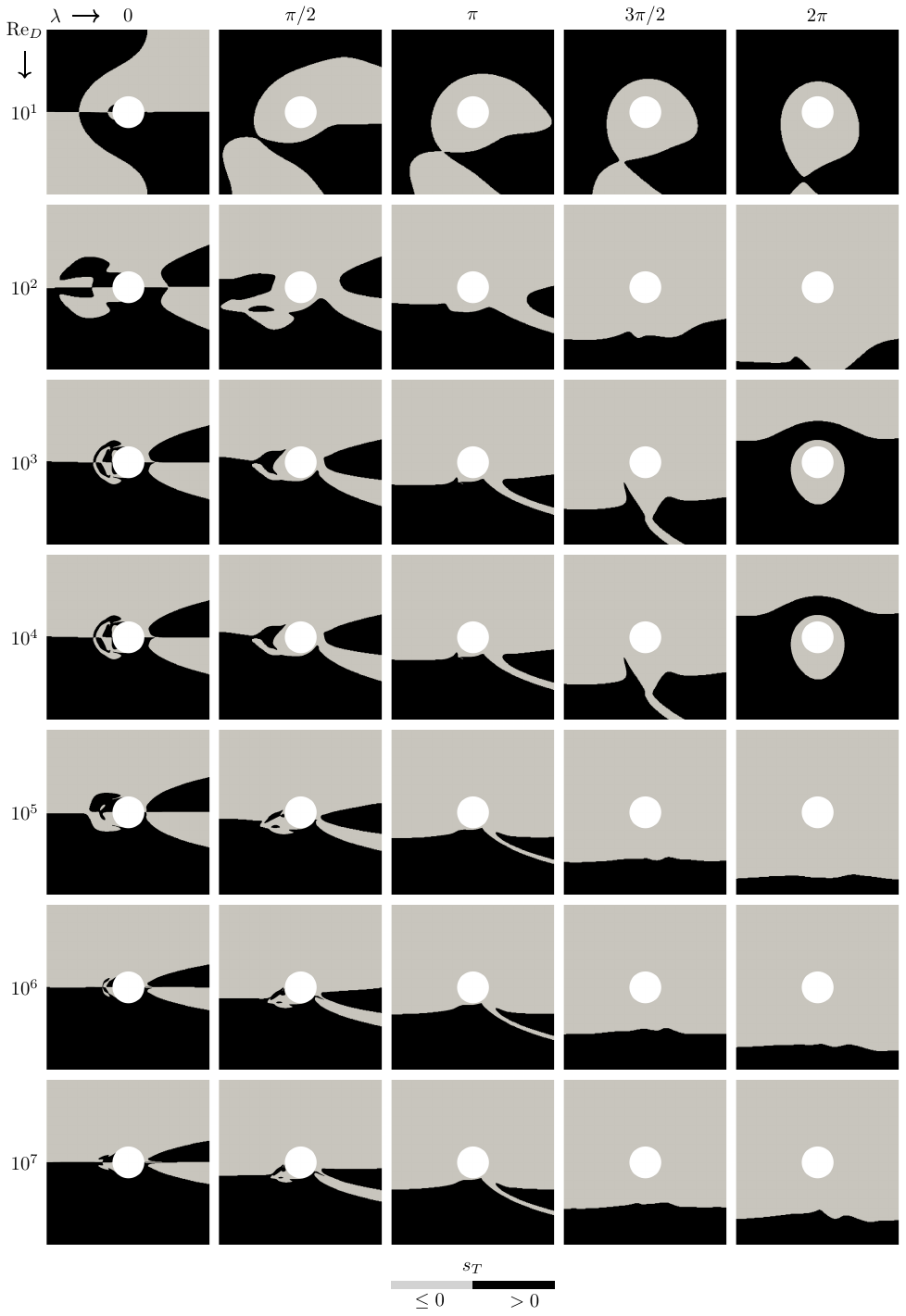}
    }
    \caption{Sign of the torque sensitivity for all investigated Reynolds numbers and spinning ratios. Black regions indicate favourable locations, whereas grey regions indicate unfavourable locations.}
    \label{fig:torque_sensitivity_spectrum}
\end{figure}

Figure~\ref{fig:torque_sensitivity_spectrum} summarizes the corresponding torque sensitivity fields. Similar to the drag and lift sensitivities, the large-scale topology depends only weakly on the Reynolds number and is governed primarily by the spinning ratio. An exception is observed for the lowest Reynolds number and, to a lesser extent, for $\mathrm{Re}_D=10^3$ at high spinning ratios, where comparatively large favourable regions are present that rapidly diminish with increasing Reynolds number. For rotating cylinders, the favourable sensitivity is predominantly located below the cylinder. With increasing spinning ratio, this favourable region continuously decreases in size, leaving only a narrow favourable band below the cylinder for the highest investigated spinning ratios.

\begin{figure}[!ht]
    \centering
    \iftoggle{tikzExternal}{
    \input{tikz/combined_sensitivity_spectrum.tikz}}{
    \includegraphics{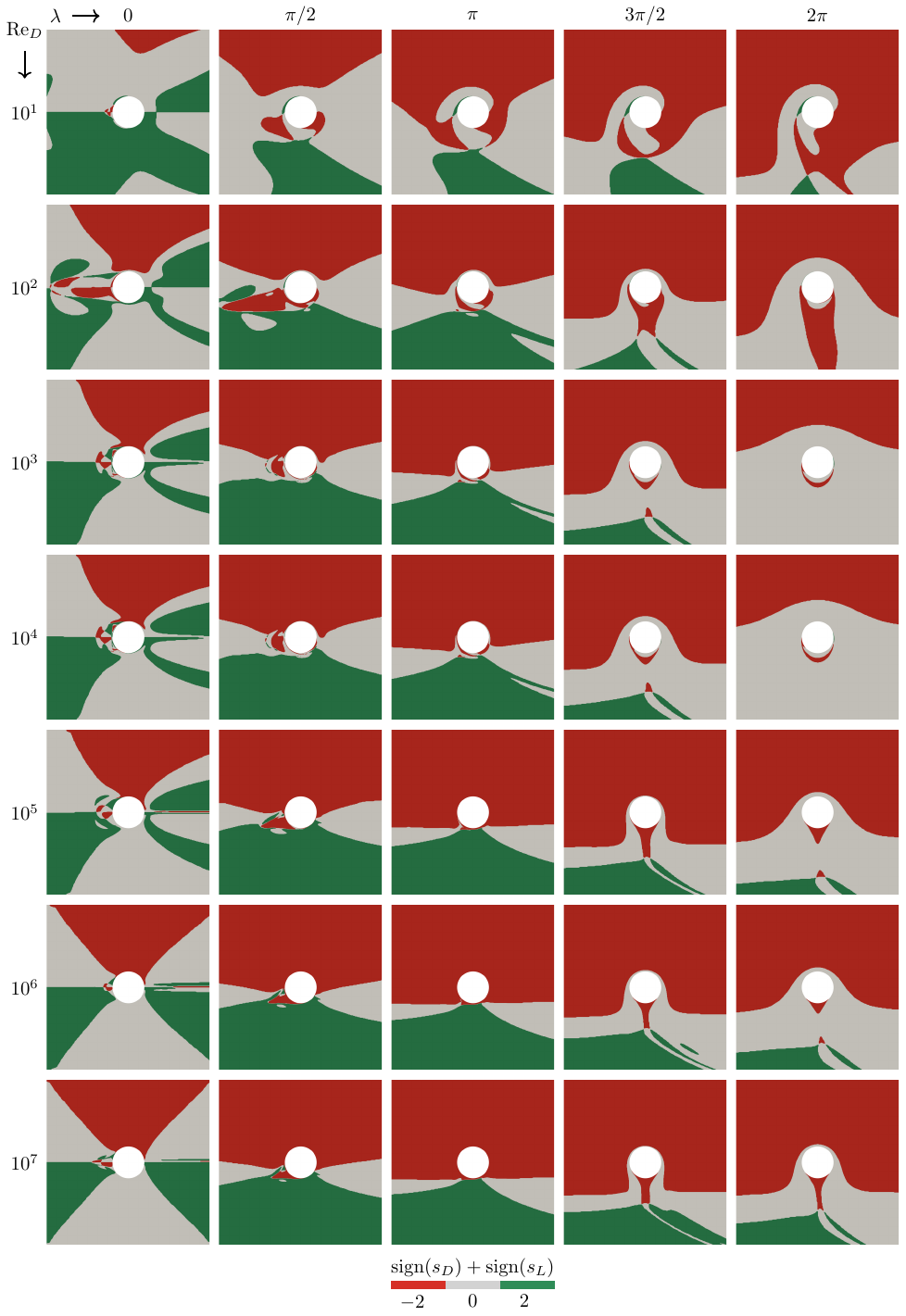}
    }
    \caption{Combined drag--lift sign fields for all investigated Reynolds numbers and spinning ratios. Green denotes favourable regions ($s_D>0$, $s_L>0$), red denotes unfavourable regions ($s_D<0$, $s_L<0$), whereas grey indicates regions with conflicting or neutral sensitivity information.}
    \label{fig:combined_sensitivity_spectrum}
\end{figure}

Figure~\ref{fig:combined_sensitivity_spectrum} presents the combined drag--lift sign fields. Similar to the individual sensitivity distributions, the combined fields depend only weakly on the Reynolds number and are governed primarily by the spinning ratio. Up to $\lambda=\pi$, favourable and unfavourable regions occupy comparable portions of the computational domain, with the favourable regions consistently located below the cylinder. For spinning ratios of $\lambda\geq3\pi/2$, however, the favourable region continuously decreases in size, whereas the unfavourable region expands and dominates large parts of the computational domain. At the same time, extended neutral regions develop in the immediate vicinity of the cylinder.

\subsection{Operating-Point Transferability}
\label{subsec:operating_point_transferability}

The previously discussed sensitivity spectra suggest that passive flow-control devices should only be transferred between operating conditions if the corresponding favourable regions remain spatially aligned. To illustrate this aspect, two additional structural validation studies are performed at $\mathrm{Re}_D=10^6$ and $\lambda=\pi$. In both cases, the forward simulations are carried out on the finest grid level of the corresponding Reynolds-number-specific grid refinement sequence in order to exclude discretization effects from the subsequent comparison.

In the first study, the passive structure originally derived from the combined drag--lift sensitivity field at $\mathrm{Re}_D=10$ is transferred unchanged to the high-Reynolds-number operating point. Since the geometry is identical to that employed in the structural validation presented in Section~\ref{subsec:structural_validation}, no additional illustration is required. As indicated by the corresponding combined sensitivity field in Fig.~\ref{fig:combined_sensitivity_spectrum}, however, the transferred structure is now located predominantly within an unfavourable region. Consequently, a deterioration of the aerodynamic performance is expected. The resulting aerodynamic coefficients are summarized in Tab.~\ref{tab:transfer_validation}.

\begin{table}[!ht]
\centering
\caption{Relative changes in the aerodynamic coefficients and lift-to-drag ratio resulting from transferring the sensitivity-informed structure from $\mathrm{Re}_D=10$ to $\mathrm{Re}_D=10^6$ at constant spinning ratio $\lambda=\pi$.}
\label{tab:transfer_validation}
\begin{tabular}{lcccc}
\toprule
Configuration &
$\Delta c_D/c_D$ [\%] &
$\Delta c_L/c_L$ [\%] &
$\Delta c_T/c_T$ [\%] &
$\Delta(c_L/c_D)/(c_L/c_D)$ [\%] \\
\midrule
Cylinder only            & +1089.539 & -48.623 & +10.604 & -95.681 \\
Cylinder + add-on        & +1267.189 & -48.607 &  +9.592 & -96.241 \\
\bottomrule
\end{tabular}
\end{table}

The transferred structure clearly fails to improve the aerodynamic performance at the new operating point. Instead, the drag increases substantially while the lift is reduced by approximately one half, resulting in a pronounced deterioration of the lift-to-drag ratio. This behaviour is fully consistent with the corresponding high-Reynolds-number sensitivity field, which classifies the occupied region as unfavourable. The result demonstrates that passive structures should not be transferred across widely separated operating conditions without re-evaluating the underlying sensitivity distribution.

\begin{figure}[!ht]
    \centering
    \iftoggle{tikzExternal}{
    \input{./tikz/structural_geometry_ReD_1E+06.tikz}}{
    \includegraphics{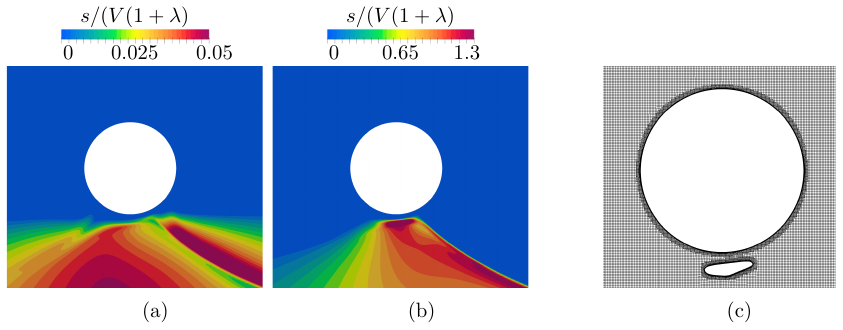}}
    \caption{Generation of the operating-point-specific passive flow-control structure at $\mathrm{Re}_D=10^6$ and $\lambda=\pi$. (a) Drag sensitivity field, (b) lift sensitivity field, and (c) body-fitted computational mesh of the resulting guide structure. The add-on geometry is obtained by extracting the $s/(V({1+\lambda}) = 1.1$ contour from the lift sensitivity field and is located entirely within a drag-reducing region.}
    \label{fig:transferability_geometry}
\end{figure}

To assess whether the sensitivity maps nevertheless provide meaningful design guidance at the higher Reynolds number, a second passive structure is generated directly from the drag and lift sensitivity fields corresponding to $\mathrm{Re}_D=10^6$ and $\lambda=\pi$. Figure~\ref{fig:transferability_geometry}(a,b) presents the corresponding drag and lift sensitivity fields, while Fig.~\ref{fig:transferability_geometry}(c) shows the resulting body-fitted computational mesh. While both objectives identify favourable regions on the decelerated side of the rotating cylinder, the lift sensitivity exhibits a particularly distinct hotspot and is therefore used to define the geometry of the guide structure. Specifically, the contour corresponding to the normalized sensitivity level $s/(V({1+\lambda}) = 1.1$ is extracted and converted into the resulting add-on geometry. The generated structure is located entirely within a region associated with drag reduction. The newly generated guide structure can clearly be identified beneath the cylinder and reflects the operating-point-specific shift of the favourable design region predicted by the sensitivity atlas. The resulting aerodynamic performance is summarized in Tab.~\ref{tab:operating_point_validation}.
\begin{table}[!ht]
\centering
\caption{Relative changes in the aerodynamic coefficients and lift-to-drag ratio resulting from an operating-point-specific sensitivity-informed structure at $\mathrm{Re}_D=10^6$ and $\lambda=\pi$.}
\label{tab:operating_point_validation}
\begin{tabular}{lcccc}
\toprule
Configuration &
$\Delta c_D/c_D$ [\%] &
$\Delta c_L/c_L$ [\%] &
$\Delta c_T/c_T$ [\%] &
$\Delta(c_L/c_D)/(c_L/c_D)$ [\%] \\
\midrule
Cylinder only            & -15.871 &  +0.262 &  -0.471 & +19.177 \\
Cylinder + add-on        &  -3.886 &  +0.126 & +15.789 &  +4.174 \\
\bottomrule
\end{tabular}
\end{table}
In contrast to the transferred geometry, the operating-point-specific structure exhibits the expected qualitative behaviour. The drag is reduced while the lift exhibits a slight increase, resulting in an improved lift-to-drag ratio for both the cylinder-only evaluation and the combined cylinder--structure system. Although the total torque increases due to the additional structure, the overall aerodynamic performance is improved. Together, both studies demonstrate that the proposed sensitivity atlas is not only suitable for identifying favourable design regions but also provides reliable guidance regarding the operating range over which passive flow-control concepts can be expected to remain effective.

\subsection{Engineering Implications}
\label{subsec:engineering_implications}

The presented sensitivity atlas provides practical guidance for the aerodynamic integration of rotating cylinders into engineering applications. In particular, Flettner rotors employed for wind-assisted ship propulsion typically operate within the range of $10^5\lesssim\mathrm{Re}_D\lesssim10^7$ and $1\lesssim\lambda\lesssim3$. Consequently, the lower three rows and the second and third columns of Fig.~\ref{fig:combined_sensitivity_spectrum} are of primary practical relevance.

Within this operating regime, the combined drag--lift sensitivity consistently identifies favourable regions on the decelerated side of the rotating cylinder. The corresponding sensitivity distributions exhibit only moderate variations with Reynolds number and spinning ratio, indicating that passive flow-control devices designed within these regions are expected to remain effective over a broad range of operating conditions rather than being restricted to a single design point.

From an operational perspective, an improved lift-to-drag ratio does not necessarily imply an increase in the absolute propulsive force generated by the rotor. However, by reducing the aerodynamic drag, passive flow-control devices may improve the net aerodynamic performance under operating conditions where the rotor would otherwise provide little or no benefit. This may allow the rotor to be deployed at lower apparent wind speeds or under less favourable operating conditions, thereby increasing its practical operating envelope.

From a practical perspective, the structural validation presented in the previous section demonstrates that the sensitivity fields can be translated directly into passive flow-control geometries. While the present study employed a manually selected sensitivity contour to generate the corresponding guide structure, this procedure could readily be generalized and automated. Rather than relying solely on the sensitivity sign, candidate regions could be identified using statistically defined sensitivity thresholds, for example based on a prescribed fraction of the maximum sensitivity or characteristic percentile values. The resulting regions could subsequently be converted into manufacturable geometries using standard geometric post-processing techniques.

Although the present study is restricted to two-dimensional flow configurations, the proposed methodology is directly applicable to three-dimensional configurations and therefore provides a promising basis for future investigations.

\section{Conclusion and Outlook}
\label{sec:conclusion_outlook}

The present work introduced a topology-based adjoint sensitivity analysis for rotating circular cylinders over a wide operating envelope, covering Reynolds numbers of $10^1\leq\mathrm{Re}_D\leq10^7$ and spinning ratios of $0\leq\lambda\leq2\pi$. Based on a porous-medium formulation, local sensitivity fields associated with the aerodynamic drag, lift, and torque objectives were derived and systematically evaluated. In addition to the local sensitivity distributions, corresponding integral sensitivity measures were introduced to characterize the overall response of the investigated operating conditions.

The predictive capability of the proposed methodology was demonstrated by three complementary validation studies. First, at $\mathrm{Re}_D=10$ and $\lambda=\pi$, sensitivity-informed Darcy-type source and sink distributions produced the expected first-order variations of the respective aerodynamic objectives. Second, at the same operating point, the combined drag--lift sensitivity field was successfully translated into a passive flow-control structure, resulting in a substantially improved lift-to-drag ratio despite the additional aerodynamic loads introduced by the guide structure. Finally, an operating-point-specific passive structure generated at $\mathrm{Re}_D=10^6$ and $\lambda=\pi$ confirmed the applicability of the proposed approach at practically relevant Reynolds numbers, whereas the unsuccessful transfer of the low-Reynolds-number geometry highlighted the importance of operating-point-specific sensitivity information.

The resulting sensitivity atlas revealed that the large-scale topology of the sensitivity fields is governed primarily by the spinning ratio, whereas the Reynolds number mainly influences the local distribution and magnitude of the sensitivities. Consequently, the proposed methodology provides not only a means of identifying favourable regions for passive flow-control devices, but also a framework for assessing the robustness of passive flow-control concepts under varying operating conditions.

Future work should extend the present methodology to fully three-dimensional configurations, including finite-length Flettner rotors, end-plate effects, and realistic ship operating conditions. Furthermore, the broad sensitivity atlas presented in this work motivates more detailed investigations of individual flow regimes, such as the laminar, transitional, and turbulent operating ranges. In addition, the sign-based combination of sensitivity fields may be generalized to multiple operating conditions, enabling the identification of passive flow-control concepts that remain effective over prescribed ranges of Reynolds number and spinning ratio. Finally, the computed topology sensitivities may serve as an initial design stage for subsequent shape optimization, thereby establishing a systematic topology-first, shape-second optimization strategy.

\section{Declaration of Competing Interest}
The author declares that he has no known competing financial interests or personal relationships that could have appeared to influence the work reported in this paper.

\section{Acknowledgments}
The current work is part of the “Propulsion Optimization of Ships and Appendages” (Grant No. 03SX599C) and "Development of a Comprehensive Methodology for the Integration of Flettner Rotors on Different Ship Types" (Grant No. 03SX581G) research projects funded by the German Federal Ministry for Economics and Climate Action. The author gratefully acknowledges this support.

\section{Acknowledgment of AI Assistance}
AI-based language tools were used to improve the wording and readability of parts of the manuscript. All scientific concepts, methodology, numerical simulations, analyses, and conclusions were developed and verified exclusively by the authors.

\section{Declaration of Competing Interest}
The authors declare that they have no known competing financial interests or personal relationships that could have appeared to influence the work reported in this paper.

\section{Data Availability Statement}
The data generated during the current study are available from the corresponding author upon reasonable request.


\bibliography{library.bib}

@article{borrvall2003topology,
  title={{Topology Optimization of Fluids in Stokes Flow}},
  author={Borrvall, T. and Petersson, J.},
  journal={International Journal for Numerical Methods in Fluids},
  volume={41},
  number={1},
  pages={77--107},
  year={2003},
  publisher={Wiley Online Library},
  doi={10.1002/fld.426}
}

@article{choi2008control,
  title={{Control of Flow Over a Bluff Body}},
  author={Choi, H. and Jeon, W.P. and Kim, J.},
  journal={Annual Review of Fluid Mechanics},
  volume={40},
  number={1},
  pages={113--139},
  year={2008},
  publisher={Annual Reviews},
  doi={10.1146/annurev.fluid.39.050905.110149}
}

@article{dilgen2018topology,
  title={{Topology Optimization of Turbulent Flows}},
  author={Dilgen, C.B. and Dilgen, S.B. and Fuhrman, D.R. and Sigmund, O. and Lazarov, B.S.},
  journal={Computer Methods in Applied Mechanics and Engineering},
  volume={331},
  pages={363--393},
  year={2018},
  publisher={Elsevier},
  doi={10.1016/j.cma.2017.11.029}
}

@article{ecca2014procedure,
  title={{A Procedure for the Estimation of the Numerical Uncertainty of CFD Calculations Based on Grid Refinement Studies}},
  author={E{\c{c}}a, L. and Hoekstra, N.},
  journal={Journal of Computational Physics},
  volume={262},
  pages={104--130},
  year={2014},
  publisher={Elsevier},
  doi={10.1016/j.jcp.2014.01.006}
}

@article{gersborg2005topology,
  title={{Topology Optimization of Channel Flow Problems}},
  author={Gersborg-Hansen, A. and Sigmund, O. and Haber, R. B.},
  journal={Structural and Multidisciplinary Optimization},
  volume={30},
  number={3},
  pages={181--192},
  year={2005},
  publisher={Springer},
  doi={10.1007/s00158-004-0508-7}
}

@article{giles2000introduction,
  title={{An Introduction to the Adjoint Approach to Design}},
  author={Giles, M.B. and Pierce, N.A.},
  journal={Flow, Turbulence and Combustion},
  volume={65},
  number={3},
  pages={393--415},
  year={2000},
  publisher={Springer},
  doi={10.1023/A:1011430410075}
}

@article{jameson1988aerodynamic,
  title={Aerodynamic {D}esign via {C}ontrol {T}heory},
  author={Jameson, A.},
  journal={Journal of Scientific Computing},
  volume={3},
  number={3},
  pages={233--260},
  year={1988},
  publisher={Springer},
  doi={10.1007/BF01061285}
}

@article{kang1999laminar,
  title={{Laminar Flow Past a Rotating Circular Cylinder}},
  author={Kang, S. and Choi, H. and Lee, S.},
  journal={Physics of Fluids},
  volume={11},
  number={11},
  pages={3312--3321},
  year={1999},
  publisher={American Institute of Physics},
  doi={10.1063/1.870190}
}

@article{kuhl2019decoupling,
  title={Decoupling of {C}ontrol and {F}orce {O}bjective in {A}djoint-{B}ased {F}luid {D}ynamic {S}hape {O}ptimization},
  author={K{\"u}hl, N. and M{\"u}ller, P. M. and St{\"u}ck, A. and Hinze, M. and Rung, T.},
  journal={AIAA journal},
  volume={57},
  number={9},
  pages={4110--4114},
  year={2019},
  publisher={American Institute of Aeronautics and Astronautics},
  doi = {10.2514/1.J058376}
}

@article{kuhl2021adjoint_2,
  title={Adjoint {C}omplement to the {U}niversal {M}omentum {L}aw of the {W}all},
  author={K{\"u}hl, N. and M{\"u}ller, P. M. and Rung, T.},
  journal={Flow, Turbulence and Combustion},
  year={2021},
  publisher={Springer},
  doi={10.1007/s10494-021-00286-7}
}

@article{kuhl2024continuous,
  title={{On the Continuous Adjoint of Prominent Explicit Local Eddy Viscosity-based Large Eddy Simulation Approaches for Incompressible Flows}},
  author={K{\"u}hl, N.},
  journal={Flow, Turbulence and Combustion},
  volume={113},
  number={2},
  pages={293--330},
  year={2024},
  publisher={Springer},
  doi={10.1007/s10494-024-00543-5}
}

@article{kuhl2025adjoint,
  title={{Adjoint-Assisted Topology-Optimization-Inspired Analysis of Pseudo-Porous Flow Fields: Application to a Flettner Rotor}},
  author={K{\"u}hl, N.},
  journal={arXiv preprint arXiv:2505.04833},
  year={2025}
}

@article{lu2020ship,
  title={{Ship Energy Performance Study of Three Wind-Assisted Ship Propulsion Technologies Including a Parametric Study of the Flettner Rotor Technology}},
  author={Lu, R. and Ringsberg, J.W.},
  journal={Ships and offshore structures},
  volume={15},
  number={3},
  pages={249--258},
  year={2020},
  publisher={Taylor \& Francis},
  doi={10.1080/17445302.2019.1612544}
}

@article{menter2003ten,
  title={{Ten Years of Industrial Experience with the SST Turbulence Model}},
  author={Menter, F.R. and Kuntz, M. and Langtry, R.},
  journal={Turbulence, Heat and Mass Transfer},
  volume={4},
  number={1},
  pages={625--632},
  year={2003}
}

@article{mittal2003flow,
  title={{Flow Past a Rotating Cylinder}},
  author={Mittal, S. and Kumar, B.},
  journal={Journal of Fluid Mechanics},
  volume={476},
  pages={303--334},
  year={2003},
  publisher={Cambridge University Press},
  doi={10.1017/S0022112002002938 }
}

@article{othmer2008continuous,
  title={{A Continuous Adjoint Formulation for the Computation of Topological and Surface Sensitivities of Ducted Flows}},
  author={Othmer, C.},
  journal={International Journal for Numerical Methods in Fluids},
  volume={58},
  number={8},
  pages={861--877},
  year={2008},
  publisher={Wiley Online Library},
  doi={10.1002/fld.1770}
}

@article{rung2009challenges,
  title={Challenges and {P}erspectives for {M}aritime {CFD} {A}pplications},
  author={Rung, T. and W{\"o}ckner, K. and Manzke, M. and Brunswig, J. and Ulrich, C. and St{\"u}ck, A.},
  journal={Jahrbuch der Schiffbautechnischen Gesellschaft},
  volume={103},
  pages={127--39},
  year={2009}
}

@article{seifert2012review,
  title={{A Review of the Magnus Effect in Aeronautics}},
  author={Seifert, J.},
  journal={Progress in Aerospace Sciences},
  volume={55},
  pages={17--45},
  year={2012},
  publisher={Elsevier},
  doi={10.1016/j.paerosci.2012.07.001}
}

@article{soto2004adjoint,
  title={{An Adjoint-Based Design Methodology for CFD Problems}},
  author={Soto, O. and L{\"o}hner, R. and Yang, C.},
  journal={International Journal of Numerical Methods for Heat \& Fluid Flow},
  volume={14},
  number={6},
  pages={734--759},
  year={2004},
  publisher={Emerald Group Publishing Limited},
  doi={10.1108/09615530410544292}
}

@article{stojkovic2002effect,
  title={{Effect of High Rotation Rates on the Laminar Flow around a Circular Cylinder}},
  author={Stojkovi{\'c}, D. and Breuer, M. and Durst, F.},
  journal={Physics of Fluids},
  volume={14},
  number={9},
  pages={3160--3178},
  year={2002},
  publisher={American Institute of Physics}
}

@article{strykowski1990formation,
  title={{On the Formation and Suppression of Vortex ‘Shedding’at Low Reynolds Numbers}},
  author={Strykowski, P.J. and Sreenivasan, K.R.},
  journal={Journal of Fluid Mechanics},
  volume={218},
  pages={71--107},
  year={1990},
  publisher={Cambridge University Press},
  doi={10.1017/S0022112090000933 }
}

@article{stuck2013adjoint,
  title={Adjoint {C}omplement to {V}iscous {F}inite-{V}olume {P}ressure-{C}orrection {M}ethods},
  author={St{\"u}ck, A. and Rung, T.},
  journal={Journal of Computational Physics},
  volume={248},
  pages={402--419},
  year={2013},
  publisher={Elsevier},
  doi={10.1016/j.jcp.2013.01.002}
}

@article{traut2014propulsive,
  title={{Propulsive Power Contribution of a Kite and a Flettner Rotor on Selected Shipping Routes}},
  author={Traut, M. and Gilbert, P. and Walsh, C. and Bows, A. and Filippone, A. and Stansby, P. and Wood, R.},
  journal={Applied Energy},
  volume={113},
  pages={362--372},
  year={2014},
  publisher={Elsevier},
  doi={10.1016/j.apenergy.2013.07.026}
}

@article{tsutsui2002drag,
  title={{Drag Reduction of a Circular Cylinder in an Air-Stream}},
  author={Tsutsui, T. and Igarashi, T.},
  journal={Journal of Wind Engineering and Industrial Aerodynamics},
  volume={90},
  number={4-5},
  pages={527--541},
  year={2002},
  publisher={Elsevier},
  doi={10.1016/S0167-6105(01)00199-4}
}

@book{zdravkovich1997flow,
  title={{Flow around Circular Cylinders: Volume 2: Applications}},
  author={Zdravkovich, M.M.},
  volume={2},
  year={1997},
  publisher={Oxford University Press}
}

@phdthesis{manzke2018development,
  title={Development of a {S}calable {M}ethod for the {E}fficient {S}imulation of {F}lows using {Dynamic} {G}oal-{O}rientated {Local} {Grid}-{Adaption}},
  author={Manzke, M.},
  year={2018},
  school={Hamburg University of Technology}
}

@phdthesis{schubert2019analysis,
  title={Analysis of {C}oupling {T}echniques for {O}verset-{G}rid {F}inite-{V}olume {M}ethods},
  author={Schubert, S.},
  year={2019},
  school={Hamburg University of Technology}
}

@phdthesis{stuck2012adjoint,
  title={Adjoint {N}avier-{S}tokes {M}ethods for {H}ydrodynamic {S}hape {O}ptimisation},
  author={St{\"u}ck, A.},
  year={2012},
  school={Hamburg University of Technology}
}
\bibliographystyle{plainnat}


\end{document}